\documentclass[aps,showpacs,prb,twocolumn,superscriptaddress,floatfix,longbibliography]{revtex4-2}
\usepackage[pdftex]{graphicx}
\usepackage{amsbsy, amssymb, amsmath, bm, mathtools}
\usepackage{xspace}
\usepackage{float}
\usepackage{bm}
\usepackage{color}
\usepackage{tabularx}
\usepackage{booktabs}
\usepackage[section]{placeins}
\usepackage[colorlinks=true,linkcolor=blue,citecolor=blue]{hyperref}
\usepackage{url}
\usepackage{threeparttable}
\usepackage{multirow}
\usepackage{enumerate}
\usepackage{enumitem}
\usepackage{color}
\usepackage{subfigure}  
\usepackage{epstopdf}
\usepackage{bbm}
\usepackage{graphicx}
\usepackage{hyperref}
\usepackage{threeparttable}
\usepackage{amsthm}
\usepackage{mathtools}
\usepackage{color}
\usepackage{tikz}
\usepackage{float}
\usepackage{empheq}
\usepackage{algorithm}
\usepackage{algpseudocode}
\usepackage{adjustbox}
\usepackage{enumitem}
\usepackage{siunitx}
\begin{document}
\title{Two-dimensional percolation model with long-range interaction}

\author{Ziyu Liu}
\thanks{These two authors contributed equally to this work}
\affiliation{Hefei National Research Center for Physical Sciences at the Microscale and School of Physical Sciences, University of Science and Technology of China, Hefei 230026, China}

\author{Tianning Xiao}
\thanks{These two authors contributed equally to this work}
\affiliation{Hefei National Research Center for Physical Sciences at the Microscale and School of Physical Sciences, University of Science and Technology of China, Hefei 230026, China}

\author{Zhijie Fan}
\email{zfanac@ustc.edu.cn}
\affiliation{Hefei National Research Center for Physical Sciences at the Microscale and School of Physical Sciences, University of Science and Technology of China, Hefei 230026, China}
\affiliation{Hefei National Laboratory, University of Science and Technology of China, Hefei 230088, China}
\affiliation{Shanghai Research Center for Quantum Science and CAS Center for Excellence in Quantum Information and Quantum Physics, University of Science and Technology of China, Shanghai 201315, China}

\author{Youjin Deng}
\email{yjdeng@ustc.edu.cn}
\affiliation{Hefei National Research Center for Physical Sciences at the Microscale and School of Physical Sciences, University of Science and Technology of China, Hefei 230026, China}
\affiliation{Hefei National Laboratory, University of Science and Technology of China, Hefei 230088, China}
\affiliation{Shanghai Research Center for Quantum Science and CAS Center for Excellence in Quantum Information and Quantum Physics, University of Science and Technology of China, Shanghai 201315, China}

\begin{abstract}
We perform large-scale simulations of the two-dimensional long-range bond percolation model with algebraically decaying percolation probabilities $\sim 1/r^{2+\sigma}$, using both conventional ensemble and event-based ensemble methods for system sizes up to $L=16384$. We accurately determine the critical points, the universal values of several dimensionless quantities, and the corresponding critical exponents. Our results provide compelling evidence that the system undergoes a crossover from short-range to long-range universality at $\sigma = 2$, in contradiction to Sak’s criterion. Notably, we observe a pronounced jump in the universal values and critical exponents at $\sigma = 2$, a feature absent from previous studies.
\end{abstract}
\maketitle

\section{Introduction}
\label{sec:intro}

Long-range (LR) interactions with algebraic decay are ubiquitous in physics, appearing in systems ranging from gravitation and electrostatics to dipolar media and complex networks~\cite{campa2014}. Incorporating such interactions can qualitatively alter the physical properties of the short-range (SR) systems, giving rise to exotic critical phenomena~\cite{dysonExistencePhasetransitionOnedimensional1969, maghrebi2017}.
Specifically, the universality of the transition in a $d$-dimensional system with coupling $J(r)\sim r^{-(d+\sigma)}$ shifts from the SR type to the LR universality when $\sigma$ drops below a threshold value $\sigma_*$.


The question of where the LR-SR crossover occurs has been at the center of a long-standing debate, particularly in the context of $O(n)$ spin models, which provide a canonical framework for addressing universality under LR couplings. Early renormalization-group (RG) analyses by Fisher \textit{et al.} indicated that, for $d/2 < \sigma < 2$, the critical behavior is controlled by the LR fixed point, with the anomalous dimension given by $\eta = 2 - \sigma$. This regime is commonly referred to as the LR, or non-classical regime. The SR behavior is recovered only for $\sigma > 2$, yielding the threshold $\sigma_* = 2$~\cite{fisher1972}. Sak later argued that anomalous dimension corrections shift this boundary to $\sigma_*=2-\eta_{\mathrm{SR}}$, yielding $\eta=\max(2-\sigma,\eta_{\mathrm{SR}})$, where $\eta_{\mathrm{SR}}$ is the anomalous dimension of the SR system~\cite{sak1973}. This scenario, known as Sak's criterion, has become widely recognized and supported by various theoretical and numerical studies~\cite{Honkonen_1989, Honkonen_1990, defenu2015, Luijten2002, angelini2014, horita2017, PhysRevE.110.064106}. Nonetheless, subsequent numerical studies by Picco \textit{et al.}~\cite{picco2012, blanchard2013} suggested a different picture, in which $\eta$ varies smoothly from the mean-field value at $\sigma = d/2$ to the SR value $\eta_{\mathrm{SR}}$ at $\sigma = 2$, implying that the threshold remains at $\sigma_* = 2$.

Recent numerical studies of the two-dimensional (2D) LR XY ($O(2)$) model~\cite{xiao_two-dimensional_2024,yao2025nonclassicalregimetwodimensionallongrange} revealed an abrupt change in transition type at $\sigma = \sigma_* = 2$, from a Berezinskii–Kosterlitz–Thouless (BKT) transition in the SR regime ($\sigma > 2$) to a conventional continuous transition in the LR regime ($\sigma \le 2$). The essential observation is that the boundary point $\sigma_*$ already belongs to the LR universality class. This motivates a \textit{fourth scenario} for LR-SR crossover: critical exponents vary smoothly within the LR regime, yet the crossover at $\sigma_*$ can be discontinuous. The reasoning is straightforward. (i) For $\sigma>\sigma_*$ the RG flow is attracted to the SR fixed point, so as $\sigma\to\sigma_*^+$ all critical exponents remain at their SR values. (ii) If the LR fixed point governs the criticality at $\sigma_*$, then at least some critical exponents $x_{\mathcal{O}}$ at this point must deviate from their SR values. This manifests a discontinuity at $\sigma_*$:
\begin{align}
x_{\mathcal{O}}(\sigma_*) \neq \lim_{\sigma \to \sigma^+_*}x_{\mathcal{O}}(\sigma),
\end{align}
Crucially, this argument would apply equally if both sides of $\sigma_*$ exhibited conventional second-order transitions. By contrast, if $\sigma_*$ lies within the SR regime, the discontinuity does not necessarily occur. Such discontinuity is not pathological but rather a legitimate signature of universality-class change. For instance, the 2D Blume-Capel model exhibits such a jump in critical exponents near its tri-critical point~\cite{blumeTheoryFirstOrderMagnetic1966,capelPossibilityFirstorderPhase1966,blumeIsingModelTransition1971}. To date, however, a discontinuous LR-SR crossover has not yet been conclusively established in LR $O(n)$ models.

The percolation model~\cite{stauffer_introduction_2018} provides an ideal framework to investigate this problem. Despite its geometric simplicity, it is closely related to $O(n)$ spin models through the random-cluster representation~\cite{fortuin1972random}. Dynamically, it can also be recast as a generalized epidemic process (GEP), which offers a unifying field-theoretical description of directed and isotropic percolation, with ordinary percolation emerging as the static limit~\cite{janssenLevyflightSpreadingEpidemic1999}. Extending the GEP to LR interactions naturally incorporates LR percolation, where the probability of infection  decays as $1/r^{d+\sigma}$. In this formulation, much like in LR $O(n)$ models, the propagator in momentum space exhibits a competition between the SR Laplacian term $\sim q^2$ and the LR Lévy-flight contribution $\sim q^\sigma$, raising a similar debate on the threshold $\sigma_*$ that separates LR and SR universality~\cite{jacobsenTransferMatrixComputation2013, linderLongrangeEpidemicSpreading2008}. Field theoretical RG analyses by Janssen \textit{et al.}~\cite{janssenLevyflightSpreadingEpidemic1999} predicted $\sigma_* = 2$, implying that SR behavior sets in only for $\sigma > 2$. In contrast, Linder \textit{et al.}~\cite{linderLongrangeEpidemicSpreading2008} reported numerical evidence for a lower boundary at $\sigma=43/24$, consistent with Sak’s scenario of a shifted crossover. Further simulations by Grassberger~\cite{grassbergerTwoDimensionalSIREpidemics2013} on 2D LR susceptible–infected–removed (SIR) epidemic processes showed that critical exponents deviate from SR values whenever $\sigma < 2$, supporting the existence of a genuine LR regime up to this boundary. These conflicting results highlight that, just as in LR $O(n)$ models, the location of $\sigma_*$ in LR percolation remains controversial, with theoretical predictions and numerical evidence not yet fully reconciled.

Beyond its theoretical relevance, percolation offers several practical advantages. Firstly, the percolation configuration can be generated directly without using Markov Chain Monte Carlo, thereby eliminating the need for thermalization and enabling simulations of extremely large systems. Secondly, in addition to conventional critical properties, the phase transition in percolation models can also be probed through a wide range of geometrical observables, such as cluster size distributions, wrapping probabilities, and shortest-path fractal dimensions~\cite{grassberger_numerical_1992,grassberger_spreading_1992,xu_critical_2021,zhou_shortest-path_2012}. These quantities not only capture the intrinsic fractal geometry of critical clusters but also allow the construction of universal dimensionless ratios that are largely free of nonuniversal factors and exhibit reduced finite-size corrections, making them highly sensitive probes of universality changes. Furthermore, a recently developed event-based ensemble method~\cite{fan_universal_2020,liExplosivePercolationObeys2023,li_explosive_2024,shi_universality_2025}, which represents the percolation process as a dynamic bond or site insertion process, provides additional advantages. By identifying a dynamic pseudo-critical point within each realization, this method allows the extraction of critical exponents without requiring precise prior knowledge of the critical threshold. It also ensures clean finite-size scaling behavior, with pseudo-critical fluctuations obeying a central-limit-type Gaussian distribution, thereby offering statistically robust access to universal quantities~\cite{liExplosivePercolationObeys2023, li_explosive_2024}. Taken together, these features make the LR percolation model particularly suitable for investigating the LR–SR crossover.

\begin{figure}[t]
    \centering
    \includegraphics[width=1\linewidth]{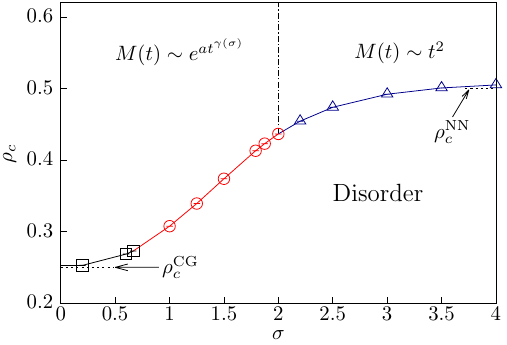}
    \caption{Phase diagram of the 2D LR interaction percolation model with probability $p\propto 1/r^{2+\sigma}$. Along the critical line, the system undergoes a continuous transition from the disordered phase below $\rho_c$ to the long-range ordered (supercritical) phase above $\rho_c$. For $\sigma > 2$, the critical behavior belongs to the 2D percolation universality class, while for $2/3 < \sigma \le 2$ the system is in the LR regime, and for $\sigma < 2/3$ it is governed by Gaussian mean-field theory. In the supercritical phase, previous studies of the LR GEP~\cite{grassberger_sir_2013, biskup_scaling_2004} have shown that for $\sigma > 2$ the cluster mass grows quadratically with time, whereas for $0 < \sigma \le 2$ it follows a stretched exponential in time. In the two limiting cases, $\lim_{\sigma\to-2}\rho_c=\rho_c^{\mathrm{CG}}=0.25$ and $\lim_{\sigma\to\infty}\rho_c=\rho_c^{\mathrm{NN}}=0.5$. 
    }
    \label{fig:PDCP}
\end{figure}

In this work, we perform large-scale simulations of the 2D LR bond percolation model with algebraically decaying percolation probability $\sim 1/r^{2+\sigma}$. The system is studied using both conventional ensemble and event-based ensemble, with the latter reaching linear sizes up to $L=16384$. We first determine the critical points with high precision in both ensembles. We then estimate the universal values of two dimensionless quantities: the critical polynomial $R_p$~\cite{xu_critical_2021} and the Binder cumulant $Q_m$. For $\sigma \leq 2$, these universal values deviate from those of the short-range regime ($\sigma > 2$), indicating that the crossover point is $\sigma_* = 2$. The phase diagram is presented in Fig.~\ref{fig:PDCP}. Notably, a discontinuity in the universal values is observed at $\sigma = 2$, a feature not reported in earlier studies. We also determine the associated critical exponents $\eta$, $\nu$, and $d_{\mathrm{min}}$, which support a consistent scenario involving a jump at $\sigma = 2$. These results provide strong evidence of a crossover from SR to LR universality at $\sigma = 2$.

The remainder of this paper is structured as follows. Section~\ref{sec:LRmodel} provides a brief introduction to the LR percolation model. Section~\ref{sec:SimObs} details the Monte Carlo algorithms employed and the two ensembles used in our simulations. Section~\ref{sec:results} presents the main numerical results. Finally, Section~\ref{sec:conclusion} summarizes the conclusions.

\section{Long-Range Percolation Model}
\label{sec:LRmodel}

We study the LR bond percolation on a $L\times L$ square lattice with periodic boundary conditions (PBC). In this model, the probability that two sites $i$ and $j$ are connected by a bond decays with their separation $r_{ij}$ as
\begin{align}
    p\left(r_{ij} \right) = K \cdot \frac{1}{r_{ij}^{2+\sigma}},
\end{align}
where $K\in(0, 1]$ is a control parameter analogous to the reduced coupling strength in $O(n)$ spin models.
Each lattice site can potentially connect to any of the remaining $N-1$ sites, with $N=L^2$ being the total number of sites. Under PBC, the square lattice is topologically equivalent to a torus, and bonds follow the shortest path on its surface, corresponding to the minimum-image convention~\cite{frenkel_understanding_2002,christiansen_phase_2019, agrawal_kinetics_2021}.

Following Ref.~\cite{xiao_two-dimensional_2024, yao2025nonclassicalregimetwodimensionallongrange}, we normalize the percolation probability with a factor $C(\sigma,L)$,
\begin{align}
    p\left( r_{ij} \right) =\rho \cdot \frac{C\left( \sigma ,L \right)}{r_{ij}^{2+\sigma}},
    \label{eq:normalized_percolation_prob}
\end{align}
such that $\sum_{j}{C\left( \sigma ,L \right)/{{r_{ij}}^{2+\sigma}}}=4$. This normalization recasts the parameter $K$ to a physically meaningful bond density $\rho =K/C\left(\sigma, L \right)$. As we demonstrate in the following sections, reformulating the model in terms of $\rho$ simplifies both theoretical analysis and numerical implementation. Moreover, this form interpolates smoothly between the $\sigma\to\infty$ and $\sigma\to-2$ limits, corresponding to 2D nearest-neighbor (NN) bond percolation and the mean-field limit of bond percolation on a complete graph (CG), respectively.

In the SR limit, $\sigma\to\infty$, the system reduces to the 2D SR bond percolation with NN connections. As the bond density $\rho$ increases, the model undergoes a continuous phase transition from a disordered state to a supercritical percolation phase, where one cluster dominates the lattice. Several critical exponents in this limit are exactly solvable~\cite{stauffer_introduction_2018}, with the anomalous dimension $\eta = 5/24$, and the correlation length critical exponent $\nu = 4/3$. 
In addition, geometric critical exponents, which characterize the fractal geometry of clusters at criticality, have been accurately determined numerically, such as the shortest-path exponent $d_{\min} = 1.130 77(2)$~\cite{PhysRevE.86.061101}.
In the $\sigma \to -2$ limit, the model reduces to a bond percolation on a CG. In this case, the critical percolation density is $1/N$, due to the normalization employed in this study, which yields $\rho_c = 1/4$. In the CG limit, the critical properties are governed by the Gaussian fixed point.

Beyond the above two limits, the critical behavior along the transition line can be divided into three regimes. (i) A mean-field regime for $\sigma\le d/3$, where the transition is governed by the Gaussian fixed point with classical mean-field exponents. (ii) An LR regime for $d/3<\sigma<\sigma_*$, governed by an LR fixed point with critical exponents varying continuously with $\sigma$. (iii) An SR regime for $\sigma>\sigma_*$, where SR exponents are recovered. Field-theoretic RG analysis in Ref.~\cite{linderLongrangeEpidemicSpreading2008} predicted a crossover at $\sigma_* = 2-\eta_{\rm SR} = 43/24$, in analogy with Sak's criterion for $O(n)$ model. Whether the SR sector extends down to $\sigma_*=2-\eta_{\rm SR}$ or only to $\sigma=2$ remains a subject of debate: Large-scale 2D SIR simulations report a singular change at $\sigma=2$ with critical exponents different from SR for all $\sigma<2$~\cite{grassbergerTwoDimensionalSIREpidemics2013}.

The supercritical phase of LR GEP in 2D has also been extensively studied both theoretically~\cite{janssenLevyflightSpreadingEpidemic1999, linderLongrangeEpidemicSpreading2008, biskup_scaling_2004, biskup_graph_2011} and numerically~\cite{grassbergerTwoDimensionalSIREpidemics2013}. The supercritical cluster mass (i.e., the number of sites in a cluster) exhibits exponential growth as a function of time $t$ in the mean-field regime $(\sigma<0)$, a stretched-exponential growth law for $0<\sigma<2$, as rigorously established by Biskup~\cite{biskup_graph_2011, biskup_scaling_2004} and numerically verified by Grassberger, and finally a power-law growth $M(t)\sim t^2$ in the SR regime $(\sigma>2)$. 
At the marginal point $\sigma = 2$, $M(t)$ follows a power-law growth, with exponents that vary continuously as a function of the percolation probability~\cite{grassbergerTwoDimensionalSIREpidemics2013}.

The correspondence between epidemic spreading and bond percolation allows one to reinterpret these results: the spreading time $t$ in GEP can be mapped to the breadth–first search (BFS) depth $s$ in percolation clusters~\cite{zhou_shortest-path_2012}. In finite-size simulations, the largest cluster dominates in the supercritical phase, reaching a mass of order $L^2$. Consequently, in the SR regime $(\sigma>2)$, the maximal BFS depth $s_{\rm max}$ in the system should scale linearly with system size, $s_{\rm max}\sim L$. For $0<\sigma<2$, Biskup's results indicate that the maximal BFS depth should obey $s_{\rm max}\sim (\ln L)^{1/\gamma(\sigma)}$, with a $\sigma$-dependent exponent $\gamma(\sigma)$ that interpolates between $1$ (as $\sigma\to 0$, corresponding to small-world/mean-field behavior) and $0$ (as $\sigma\to 2$)~\cite{biskup_graph_2011}. Finally, for $\sigma<0$, the BFS depth becomes logarithmic in system size, $s_{\rm max}\sim \ln L$, reflecting the tree-like mean-field structure. A detailed numerical verification of the supercritical scaling forms, in particular of the stretched-exponential exponent $\gamma(\sigma)$, is highly challenging~\cite{grassbergerTwoDimensionalSIREpidemics2013} and lies beyond the scope of this work.

\section{Algorithms AND OBSERVABLES}
\label{sec:SimObs}

In this section, we present the two algorithms employed to simulate the LR percolation model, namely the conventional ensemble method and the event-based ensemble method, along with the measurement procedures used in this study.

\subsection{Conventional Ensemble}


In the conventional ensemble (CE), configurations of bond percolation are sampled at a fixed bond density $\rho$. By choosing $\rho=\rho_c$, one can measure critical observables and obtain their averages over different realizations. A straightforward way to generate bond percolation is to examine each possible bond $b$ on the lattice sequentially and activate it with the corresponding probability $p_b$. For the LR percolation model defined by Eq.~\eqref{eq:normalized_percolation_prob}, the total number of bonds grows as $O(N^2)$. As a result, this naive approach has a time complexity of $O(N^2)$ to generate an LR percolation configuration, which severely limits large–scale simulations and hinders precise studies of critical behavior.

An efficient way to overcome this $O(N^2)$ bottleneck in LR percolation simulations is to directly sample the activated bonds rather than inspecting all possible ones~\cite{linderLongrangeEpidemicSpreading2008, grassbergerTwoDimensionalSIREpidemics2013}. The key observation is that, near criticality, only $O(N)$ bonds are active in a system of $N$ sites, making exhaustive enumeration wasteful. Here, we adopt the clock sampling strategy, inspired by the clock Monte Carlo method~\cite{michel2019}, which generates the ``next bond activation event” directly~\cite{xiao_two-dimensional_2024, yao2025nonclassicalregimetwodimensionallongrange}. The procedure is very similar to the cluster generation strategy of the cluster algorithm for the $O(n)$ spin model detailed in Ref.~\cite{yao2025nonclassicalregimetwodimensionallongrange}, except that in the percolation case, the resampling step is unnecessary, because every sampled bond will be activated.

In the LR percolation model, there are $(N-1)/2$ distinct displacement vectors $\vec{r}$. For each $\vec{r}$, translational invariance under PBC ensures $N$ equivalent bonds $b=(i,\vec{r})$, resulting in a total of $N(N-1)/2$ candidates~\cite{luijten1995}. All bonds belonging to the same displacement type then share an identical activation probability $p(\vec{r})$ given by Eq.~\eqref{eq:normalized_percolation_prob}. To construct a configuration, we treat each displacement type independently. Instead of examining all $N$ bonds sequentially, we directly sample the position of the next activated bond. The key idea is that bond activations within a given type can be regarded as a sequence of independent Bernoulli trials with a success probability $p$. The probability that the first $k$ bonds are inactive and the $(k+1)$-th is active follows a geometric distribution, and from its cumulative form, one obtains the update rule for the bond index:
\begin{align}
k \leftarrow k + 1 + \left\lfloor\frac{\ln(1-u)}{\ln(1-p)}\right\rfloor,
\end{align}
where $u$ is a uniform random number in $[0,1)$ and $\left\lfloor \cdot \right\rfloor$ denotes the floor function. If $k \le N$, the $k$-th bond is marked as active; otherwise, the process for this displacement type terminates. The procedure is then repeated until all displacement types have been processed. In this way, a complete set of activated LR bonds is obtained. Compared with the naive $O(N^2)$ enumeration, this method requires only $O(N)$ operations, enabling efficient simulations of large systems while preserving the exact activation statistics in Eq.~\eqref{eq:normalized_percolation_prob}.

For each percolation configuration, the following quantities are measured:
\begin{enumerate}[%
  label=(\alph*),        
  leftmargin=2em,        
  labelwidth=1.5em,      
  labelsep=0.5em,        
  align=left             
]

\item The size of the largest cluster $\mathcal{C}_1$.
\item The second and fourth moments of the cluster‐size distribution, $\mathcal{S}_2=\sum_i \mathcal{C}_i^2$ and $\mathcal{S}_4=\sum_i \mathcal{C}_i^4$.
\item The wrapping indicators $\mathcal{R}_2$ and $\mathcal{R}_0$: $\mathcal{R}_2 = 1$ if a cluster wraps around the system in both directions, and $\mathcal{R}_2 = 0$ otherwise. Similarly, $\mathcal{R}_0 = 1$ if no clusters exhibit wrapping, and $\mathcal{R}_0 = 0$ otherwise.
\item The Fourier mode of the magnetization $\mathcal{M}_{\mathbf{k}}=\bigl|\sum_i s_i\,e^{i\,\mathbf{k}\cdot\mathbf{r}_i}\bigr|$, obtained by assigning a random sign $s_i\in\{\pm 1\}$ to each percolation cluster, with $\mathbf{k}=(2\pi/L,0)$.


\item The longest graph distance $\mathcal{S}_1$ is determined as follows. For each cluster, we perform a BFS from a seed site and denote the maximum depth $\mathcal{S}$ as the graph distance of the cluster. We then sample the longest graph distance $\mathcal{S}_1$ among all clusters.

\end{enumerate}

From these quantities, we sample the following observables:
\begin{enumerate}[%
  label=(\alph*),        
  leftmargin=2em,
  labelwidth=1.5em,
  labelsep=0.5em,
  align=left
]
\item The averaged largest‐cluster size $C_1=\langle\mathcal{C}_1\rangle$.
\item The susceptibility $S_2=\langle\mathcal{S}_2\rangle$.
\item The Fourier‐mode susceptibility $\chi_k=L^{-2}\langle\mathcal{M}_{\mathbf{k}}^2\rangle$.
\item The fourth‐order Binder cumulant 
      $Q_m=\langle\mathcal{S}_2\rangle^2/\bigl(3\langle\mathcal{S}_2^2\rangle-2\langle\mathcal{S}_4\rangle\bigr)$.
\item The critical polynomial $R_p=\langle\mathcal{R}_2-\mathcal{R}_0\rangle$.
\item The averaged maximal distance $S_1 = \left< \mathcal{S}_1\right>$
\end{enumerate}

\subsection{Event-Based Ensemble}
\label{event-based}

The event‐based ensemble (EB) method was originally introduced to address the anomalous fractal dimension observed in explosive percolation~\cite{liExplosivePercolationObeys2023} and has since been widely applied to various percolation models~\cite{lu2024selfsimilargapdynamicspercolation,shi_universality_2025}. The key idea is to interpret percolation as a dynamic bond-insertion process, in which the critical behavior of observables is extracted from the statistics of pseudo‐critical events. Here, a pseudo‐critical event is defined as the step at which the largest cluster undergoes its most abrupt growth upon the insertion of a new bond.

To simulate the LR bond percolation, we first generate a set of candidate long‐range bonds $\mathcal{B}$ based on Eq.~\eqref{eq:normalized_percolation_prob} using the aforementioned clock sampling method. Starting from an empty system at time $t = 0$, at each subsequent step a bond is randomly drawn from $\mathcal{B}$, and inserted into the system, and $t$ is incremented by $1$, until $\mathcal{B}$ is empty. During the process, we record the size of the largest cluster $\mathcal{C}_1(t)$ and compute the gap value $\delta(t) = \mathcal{C}_1(t) - \mathcal{C}_1(t-1)$, which quantifies the growth of the largest cluster due to the newly inserted bond. A pseudo‐critical event is identified when $\delta(t)$ attains its maximum over the entire process, $\delta(t_{1}) = \max_{0 < t \le |\mathcal{B}|}\delta(t)$. The corresponding bond density $\varrho_1 = t_1/2N$ is defined as the first pseudo‐critical point. Analogously, the second pseudo‐critical point $\varrho_2$ is determined by the step $t_2$ where $\delta(t)$ takes its second‐largest value, with ties resolved by the earliest occurrence. Finally, using the recorded sequence of bond insertions, the percolation configuration at $\varrho_1$ and $\varrho_2$ can be reconstructed, allowing any observables of interest to be sampled. The EB ensemble average is defined as the mean of these observables measured at the dynamically determined pseudo‐critical points.

For the percolation configurations generated at the pseudo-critical points, we measure the same physical quantities as in the CE. Additionally, we record the following quantities during the bond insertion process:
\begin{enumerate}[%
  label=(\alph*),        
  leftmargin=2em,        
  labelwidth=1.5em,      
  labelsep=0.5em,        
  align=left             
]
  \item The gap value at the pseudo-critical point $\delta(t_{1})$.
  \item The position of the first and second pseudo‐critical points $\varrho_1$ and $\varrho_2$.
\end{enumerate}

From these measurements, we can compute:
\begin{enumerate}[%
  label=(\alph*),        
  leftmargin=2em,        
  labelwidth=1.5em,      
  labelsep=0.5em,        
  align=left             
]
  \item The averaged pseudo-critical gap $\Delta = \langle \delta(t_{1}) \rangle$.
  \item The averaged pseudo-critical point $\rho_1 = \langle \varrho_1 \rangle$ and $\rho_2 = \langle \varrho_2 \rangle$.
  \item The standard deviation of the first pseudo‐critical point $s(\varrho_1) = \sqrt{\langle \varrho_1^2 \rangle - \langle \varrho_1 \rangle^2}$.
  \item The critical window $\langle |\varrho_1 - \varrho_2| \rangle$.
\end{enumerate}

\section{Results}
\label{sec:results}

In this work, we focus on several representative cases near the crossover regimes, i.e., $\sigma=2.2, 2.0, 15/8$, and $43/24$. For these values, large-scale simulations are performed with system sizes up to $L=4096$ in CE and $L=16384$ in EB, and more than $10^5$ samples for each data point. Statistical errors were estimated using the binning method for simple observables and the Jackknife method for composite ones. From the combined CE and EB data, we determine the critical point $\rho_c$, the universal ratios of the critical polynomial $R_p$ and the Binder cumulant $Q_m$, as well as several critical exponents: the anomalous dimension $\eta$, correlation length exponent $\nu$ and the shortest-path exponent $d_{\mathrm{min}}$~\cite{PhysRevE.86.061101}. The main numerical results are summarized in Fig.\ref{fig:PDCP}, Table~\ref{table:critical_points}, and Table~\ref{table:universality}.

\begin{table}[h]
\centering
\caption{Summary of critical points $\rho_c$ for various $\sigma$. High accuracy results are obtained for $\sigma=2.2, 2.0, 15/8$, and $43/24$. For other $\sigma$ values, where the finite-size effects are much weaker, simple estimates are given for illustration. }
\begin{adjustbox}{max width=\textwidth}
\begin{tabular}{ll|ll}
\hline\hline
\multicolumn{1}{c}{$\sigma$} &\multicolumn{1}{c}{$\rho_c$}     & \multicolumn{1}{c}{$\sigma$} & \multicolumn{1}{c}{$\rho_c$}  \\ \hline
NN       & 0.5          & 3/2   & 0.373 96(2)  \\
2.2      & 0.454 409(1) & 5/4  & 0.339 35(4)  \\
2.0      & 0.436 505(1) & 1.0   & 0.307 63(8)  \\
15/8     & 0.423 003(2) & 3/5   & 0.269 0(2)   \\
43/24    & 0.413 053(2) & -2    & 0.25         \\ \hline\hline
\end{tabular}
\end{adjustbox}
\normalsize
\label{table:critical_points}
\end{table}

\begin{table}[h]
\centering
\caption{Summary of the universal values and critical exponents. The Binder cumulant for the NN case is cited from Ref.~\cite{PhysRevB.71.144303} and $d_{\rm min}$ from Ref.~\cite{zhou_shortest-path_2012}}
\begin{adjustbox}{max width=\textwidth}
\begin{tabular}{l|lllll}
\hline\hline
$\sigma$ & \multicolumn{1}{c}{$R_p$}     & \multicolumn{1}{c}{$Q_m$}      & \multicolumn{1}{c}{$\eta$}      & \multicolumn{1}{c}{$y_t\equiv1/\nu$} & \multicolumn{1}{c}{$d_{\mathrm{min}}$}\\ \hline
NN       &\, 0         & 0.870 48(5)             & 5/24        & 3/4          & 1.130 77(2) \\
2.2      &\, 0.000(1)  & 0.869 5(10)             & 0.210(2)    & 0.750(1)     & 1.130(1)  \\ \hline
2.0      & -0.024(4)   & 0.861(2)                & 0.224(2)    & 0.743(2)     & 1.108(2)   \\
15/8    & -0.076(6)    & 0.842(3)                & 0.248(2)    & 0.733(2)     & 1.073(3)   \\
43/24   & -0.143(7)   &  0.821(4)                & 0.281(2)    & 0.723(2)     & 1.058(3)   \\ \hline\hline
\end{tabular}
\end{adjustbox}
\normalsize
\label{table:universality}
\end{table}

\subsection{Critical points}
\label{subsec:criticalpoint}

\begin{figure*}[ht]
    \centering
    \includegraphics[width=\linewidth]{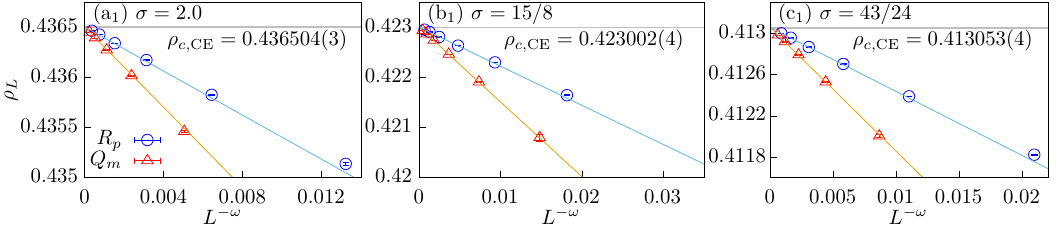}
    \includegraphics[width=\linewidth]{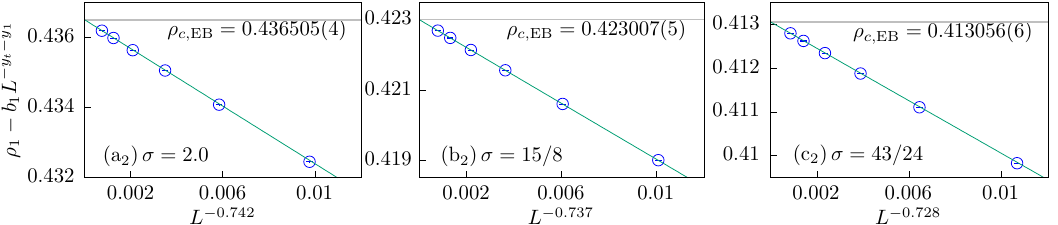}
    \vspace*{-6mm}
    \caption{Extrapolation of $\rho_L$ in the CE and $\rho_1$ in the EB. Panels (a$_1$)–(c$1$) show $\rho_L$, obtained from $R_p$ and $Q_m$ in the CE, plotted against $L^{-\omega}$ for $\sigma = 2.0, 15/8,$ and $43/24$. The correction-to-scaling exponent $\omega$ is extracted by fitting Eq.~\eqref{eq:rho_L}, yielding $\omega(R_p)=1.04,\omega(Q_m)=1.09$ for $\sigma=2.0$; $\omega(R_p)=0.964,\omega(Q_m)=1.013$ for $\sigma=15/8$; and $\omega(R_p)=0.93,\omega(Q_m)=0.98$ for $\sigma=43/24$. In all cases, $\rho_L$ from both observables converges to straight lines, leading to a consistent estimate of the critical point $\rho_{c,\mathrm{CE}}$ in the $L\to\infty$ limit, indicated by the gray band whose width represents the uncertainty. Panels (a$_2$)–(c$_2$) present the extrapolation of $\rho_1$ in the EB as a function of $L^{-y_t}$, with the leading correction term subtracted to highlight its convergence. The critical points inferred from both ensembles agree within error bars and are consistent with the FSS least-squares fits.}   

    \label{fig:LR_Rp_Qm_Perco_ex_x0}
\end{figure*}

We first determine the critical point $\rho_c$ for various values of $\sigma$ and map out the overall phase diagram of the system. Critical points are obtained from both the CE and the EB, and the two methods yield consistent results within uncertainty. Two complementary strategies are employed: finite-size scaling (FSS) analyses of the dimensionless ratios $R_p$ and $Q_m$, and extrapolations of pseudo-critical points $\rho_L$.

\textit{FSS in CE} --- The critical points are first estimated using FSS analysis of the critical polynomial $R_p$ and Binder cumulant $Q_m$ obtained in CE. These dimensionless quantities exhibit universal values at $\rho_c$, and in the critical scaling regime of a second-order transition, they satisfy the finite-size scaling relation,
\begin{align}
    \mathcal{O}(t,u,L)=\mathcal{O}(tL^{y_t},uL^{y_1},1)+\cdots,
    \label{eq:Qm_scale}
\end{align}
where $t\equiv(\rho-\rho_c)/\rho_c$ is the distance to the critical point, $y_t$ is the thermal scaling exponent. Here, $u$ and $y_1$ are the leading irrelevant field and its corresponding scaling exponent. The ellipsis denotes omitted correction terms arising from the field dependence of the analytic part of the free energy and other higher-order contributions.

In particular, the critical polynomial $R_p$, originally introduced by Jacobsen and Scullard in the Fortuin–Kasteleyn (FK) representation, has proven highly effective in locating critical points of 2D lattices with remarkable precision~\cite{jacobsenTransferMatrixComputation2013}. For 2D SR bond percolation, the critical polynomial has the universal value $R_p \equiv R_2 - R_0 = 0$ at $\rho_c$, which holds exactly~\cite{jacobsenTransferMatrixComputation2013}. Thus, the finite-size estimates of $\rho_c$ converge directly to the thermodynamic critical point without corrections in the SR limit.

We perform least-squares fits for these dimensionless quantities using the standard FSS ansatz of second-order transitions:
\begin{align}
\label{eq:fss_fitting_ansatz}
\mathcal{O}(\rho,L)=&q_0+\sum_{i=1}^m{q_i\left[ \left( \rho-\rho_c \right) L^{y_t} \right] ^i}+b_1L^{-y_1} \nonumber \\ 
&+b_2L^{y_2}+b_3L^{y_3}+c_1(\rho-\rho_c)L^{-y_1+y_t},
\end{align}
where $\mathcal{O} \in\{R_p,\,Q_m\}$. Here, $q_0$ is the universal value at criticality, $y_t \equiv 1/\nu$ is the thermal scaling exponent, and $y_1$ is the scaling exponent of the leading irrelevant field. The $y_2$ and $y_3$ terms represent other finite-size corrections. The final term accounts for the cross-term between the thermal and irrelevant fields. The coefficients $q_i$, $b_i$ and $c_1$ are non-universal.

For the $\sigma=2.2, 2.0, 15/8$, and $43/24$, we accurately determined the critical points down to $6$ decimals (See appendix for fitting details). Moreover, the fit also shows that the $L_{\mathrm{min}}$ required to obtain a reasonable fit for $R_p$ is smaller than that of $Q_m$. This suggests that the critical polynomial $R_p$ is less susceptible to finite-size effects than the conventional Binder ratio and could serve as a suitable quantity for examining the critical scaling behavior in this regime. We also estimate the critical points for other $\sigma$s. For $\sigma >2.2$, $R_p$ is almost free from finite-size corrections, and the critical points can be estimated directly from the crossing point of curves of different system sizes. Near and in the mean-field regime, the $Q_m$ shows fewer finite-size corrections, and critical points can be obtained from a similar fitting process. The results are summarized in Table~\ref{table:critical_points} and the phase diagram is plotted in Fig.~\ref{fig:PDCP}. As the $\sigma$ decreases to $0$, the critical point shifts from the NN value of $1/2$ to the CG limit of $1/4$.

\textit{Extrapolation in CE}---As a complementary approach, we perform separate analyses of critical points by extrapolating the pseudo-critical points $\rho_L$, defined as the intersections of curves for sizes $L$ and $2L$ [e.g., $R_p(L,\rho)$ with $R_p(2L,\rho)$]. These intersections can be obtained either by interpolation or by fitting a truncated quadratic version of Eq.~\eqref{eq:fss_fitting_ansatz}. The pseudo-critical points asymptotically converge to the critical points $\rho_c$ in the $L\rightarrow \infty$ limit as
\begin{align}
     \rho_L = \rho_c + L^{-\omega}(a+bL^{{y}_2}),
     \label{eq:rho_L}
\end{align}
where $\omega \equiv y_t + y_1$ consists of the contribution of both the thermal scaling field and the leading irrelevant field. $a$ and $b$ are non-universal coefficients, and ${y}_2$ is the finite-size correction term. The upper three panels of Fig.~\ref{fig:LR_Rp_Qm_Perco_ex_x0} illustrate the extrapolations for $\sigma=2.0,15/8,$ and $43/24$, with $\rho_L$ plotted against $L^{-\omega}$. The value of $\omega$ is specified in the caption. The extrapolation curves of both $R_p$ and $Q_m$ converge to a common $\rho_c$ in the thermodynamic limit, with uncertainties indicated by the shaded bands. Although this approach uses fewer data points and thus yields larger error bars, the extrapolated results agree with the FSS estimates, confirming consistency.

\textit{Extrapolation in EB} — In the EB simulation, the averaged pseudo-critical point $\rho_1$ is measured directly. The values of $\rho_1$ converge asymptotically to $\rho_c$ as
\begin{align}
\rho_1(L) \equiv \rho_L = \rho_c + L^{-y_t}(a+b_1L^{y_1}+b_2L^{y_2}+b_3L^{y_3}),
\end{align}
where $y_1$, $y_2$, and $y_3$ are finite-size corrections and this scaling form is fitted to obtain $\rho_c$ (see Table~\ref{fit:rho_1}). The lower panels of Fig.~\ref{fig:LR_Rp_Qm_Perco_ex_x0} show $\rho_1$ versus $L^{-y_t}$ for different $\sigma$ values, together with the fitted $\rho_c$. The EB results agree with the FSS estimates within uncertainties.

Taken together, the three approaches yield mutually consistent results, demonstrating that our estimation of $\rho_c$ is robust and reliable. 

\subsection{Dimensionless Quantities}

\begin{figure*}[t]
    \centering
    \includegraphics{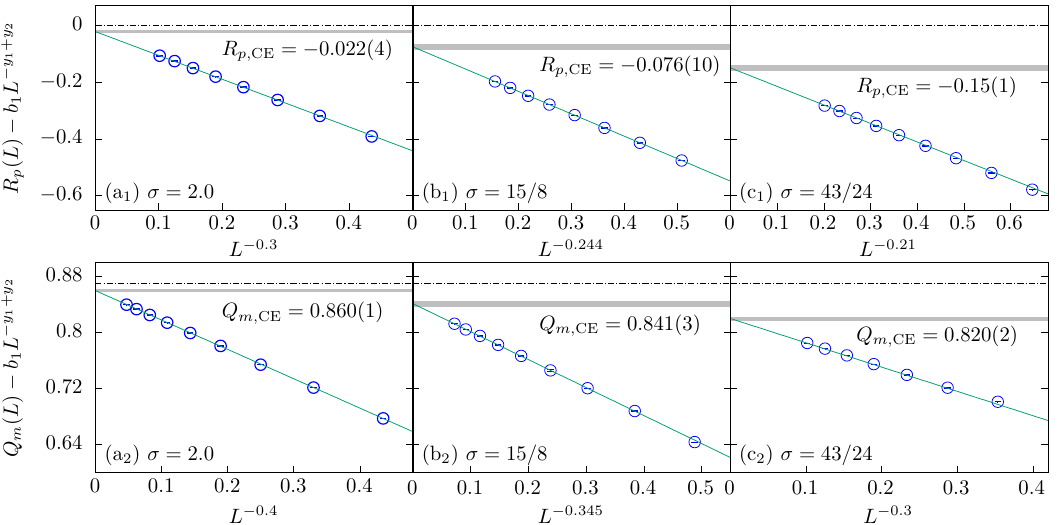}
    \vspace*{-6mm}
    \caption{Extrapolation of the universal ratios $R_p(L)$ and $Q_m(L)$ in the CE. Panels (a$_1$)–(c$_1$) display $R_p(L)$ after subtracting a finite-size correction $L^{-y_1+y_2}$, plotted as functions of $L^{y_1}$, obtained by fitting Eq.~\ref{eq:X_L}. The gray-shaded bands indicate the estimated universal values $R{p,\mathrm{CE}}$ with their associated uncertainties. Panels (a$_2$)–(c$_2$) present the corresponding extrapolation of $Q_m(L)$. In all cases, the extrapolated results are consistent with standard FSS least-squares fits within one error bar. The universal values of both $R_p$ and $Q_m$ for $\sigma \le 2$ deviate from the SR universality, marked by the dashed lines with $R_p^\mathrm{SR}=0$ and $Q_m^\mathrm{SR}=0.87048(5)$.}
    \label{fig:LR_Rp_Qm_Perco_ex_y0}
\end{figure*}

Dimensionless quantities such as $R_p$ and $Q_m$ are not only powerful tools for locating critical points, but also reliable indicators of universality. At $\rho_c$, they take universal values determined by the underlying universality class. When the universality of transition changes, these quantities also change. In the LR model, the RG flow is attracted to the SR Wilson–Fisher fixed point for $\sigma > 2$, and $R_p$ and $Q_m$ converge to SR values. By contrast, for $\sigma \leq \sigma_*$, the flow is directed to a distinct LR fixed point, leading to different universal ratios and exponents. This shift marks the boundary between the SR and LR universality classes.

From the FSS analysis in the previous section, we obtain the universal values of $Q_m$ and $R_p$, summarized in Table~\ref{table:universality}. For $\sigma=2.2$, both quantities agree with the NN case within error bars. In contrast, for $\sigma = 15/8$ and $43/24$, they exhibit explicit $\sigma$ dependence and deviate clearly from the SR values, signaling a change of universality. These results strongly suggest that the LR-SR crossover does not occur at $\sigma = 2-\eta_{\rm SR}=43/24$, but instead at $\sigma =2$, consistent with Ref.~\cite{grassbergerTwoDimensionalSIREpidemics2013}. Moreover, at $\sigma = 2$, we find that $R_p = -0.024(4)$ and $Q_m=0.861(2)$, which differ from the SR values, which indicates that the system already lies in the LR regime and flows to LR fixed points, analogous to the LR XY model~\cite{yao2025nonclassicalregimetwodimensionallongrange}.

\begin{figure}[h]
    \centering
    \includegraphics[width=1\linewidth]{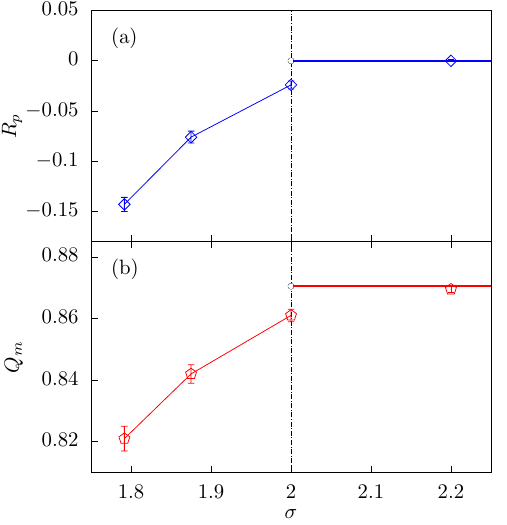}
    \vspace*{-6mm}
    \caption{Universal values of $R_p$ and $Q_m$ at $\rho_c$ as functions of $\sigma$. Panel (a) shows $R_p(\sigma)$, with the horizontal red line marking the SR value $R_p^{\text{SR}}=0$. Panel (b) shows $Q_m(\sigma)$ with the SR value $Q_m^{\text{SR}}=0.87048(5)$ marked by the horizontal blue line. At $\sigma=2.0$, both $R_p$ and $Q_m$ exhibit a discontinuous jump, deviating significantly from their SR predictions.}
    \label{fig:PD}
\end{figure}

To further confirm these findings, we also employ an extrapolation method. We carefully fit $R_p$ and $Q_m$ at their pseudo-critical points $\rho_L$ in CE, using the ansatz,
\begin{align}
  \mathcal{O}_L = \mathcal{O}_0 + L^{-y_1}(a+bL^{{y}_2}),
  \label{eq:X_L}
\end{align}
where $\mathcal{O}_0$ is the universal value, $y_1$ originates from the irrelevant field and ${y}_2$  denotes the additional finite-size correction.  Fig.~\ref{fig:LR_Rp_Qm_Perco_ex_y0} shows $R_p$ and $Q_m$ at $\rho_L$ as a function of $L^{-y_1}$ for $\sigma = 2.0, 15/8, 43/24$. The exponents $y_1$ are obtained from fitting, and the shaded gray band denotes the extrapolated limit $R_{p,\text{CE}}$ and $Q_{m,\text{CE}}$ in the $L\!\rightarrow\!\infty$ limit. In all cases, the data points asymptotically approach a straight line. Although the convergence appears slow due to sub-leading corrections, the extrapolated values remain consistent with the FSS results within one standard error. The fitted exponent $y_1$ also agrees with that obtained from the FSS analysis, supporting the robustness of our estimates. For $\sigma = 2.0, 15/8,$ and $43/24$, the universal values $R_{p,\text{CE}}$ and $Q_{m,\text{CE}}$ clearly deviate from the SR values (dotted lines), indicating a change of universality.

Finally, Fig.~\ref{fig:PD} summarizes the dependence of universal ratios on $\sigma$. For $\sigma > 2$, the $Q_m$ and $R_p$ retain their SR values, marked by a horizontal line. At $\sigma = 2$, they exhibit discontinuous jumps, identifying the boundary between SR and LR universality classes. Within the LR regime ($\sigma \le 2$), the critical points vary with $\sigma$. Taken together, these results provide strong evidence for the fourth scenario, in which $\sigma_* = 2$ and the critical properties, at least for the universal values of $R_p$ and $Q_m$, display a discontinuity at this point.

\subsection{Anomalous Dimension}





\begin{figure}[t]
    \centering
    \includegraphics[width=\linewidth]{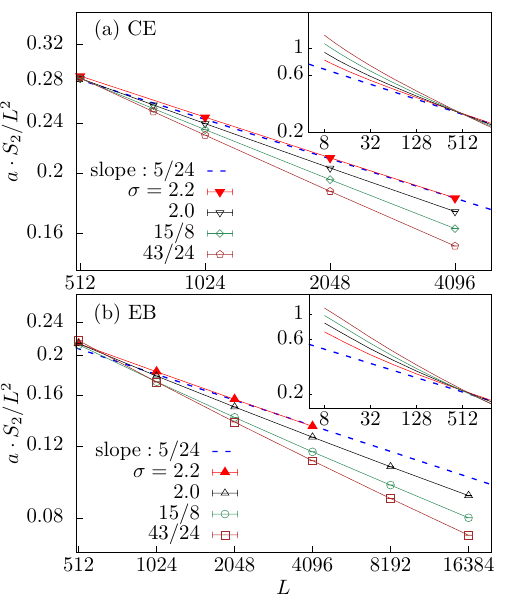}
    \vspace*{-6mm}
    \caption{Log–log plots of $S_2/L^2$ versus $L$ for various $\sigma$ in the CE (a) and EB (b). The dashed line indicates the asymptotic scaling of the NN case, with the SR exponent $5/24$. For clarity, the data are rescaled by a constant factor $a$ so that curves for different $\sigma$ coincide at $L=512$, facilitating direct comparison. The insets highlight the small-$L$ behavior in both ensembles. While finite-size corrections are pronounced at small $L$, the distinct critical exponents for different $\sigma$ remain evident at sufficiently large $L$.}
    \label{fig:S2_all}
\end{figure}

We now turn to the critical exponents of the system. The anomalous dimension $\eta$ lies at the center of the debate on $\sigma_*$. Earlier RG studies, based on the $\epsilon$-expansion to one-loop order, suggest that $\eta = 2-\sigma$ holds exactly throughout the LR regime~\cite{janssenLevyflightSpreadingEpidemic1999}. From this, it has been argued that the SR universality class prevails down to $\sigma > 2-\eta_{\rm SR}$~\cite{linderLongrangeEpidemicSpreading2008}. A similar claim was also made for the LR $\mathcal{O}(n)$ spin model~\cite{fisher1972, sak1973}. However, numerical studies of both the LR $\mathcal{O}(n)$ spin model~\cite{picco2012, yao2025nonclassicalregimetwodimensionallongrange} and the LR GEP model~\cite{grassbergerTwoDimensionalSIREpidemics2013} show that $\eta$ deviates from this prediction as $\sigma \to 2$. While strong corrections are indeed expected near $\sigma = 2$, these deviations should not be dismissed as mere finite-size or finite-time artifacts.

To resolve the value of $\eta$ in this crossover regime, we systematically analyze the critical scaling of several observables. In the CE, we examine the scaling of $C_1$, $S_2$, and $\chi_k$ at the critical point $\rho_c$. In the EB, we study the scaling of the same observables at the dynamic pseudo-critical point $\rho_L$, together with the averaged gap $\Delta$. The asymptotic scalings of $S_2$ and $\chi_k$ in both ensembles are expected to follow
\begin{align}
\mathcal{O} = L^{2 - \eta}(a + b_1 L^{y_1}+b_2L^{y_2}),
\label{eq:S2_scaling}
\end{align}
while $C_1$ and $\Delta$ follow
\begin{align}
\mathcal{O} = L^{2 - {\eta}/{2}}(a + b_1 L^{y_1}+b_2L^{y_2}),
\end{align}
where $y_1$ and $y_2$ are finite-size correction exponents; $a$, $b_1$ and $b_2$ are nonuniversal constants. In the following, we discuss $S_2$ under the EB ensemble in detail as a representative case and show that the critical exponents extracted from all other observables agree with high accuracy. 


For $S_2$ at $\sigma = 2.2$, we first fit with only one free correction $y_1$ term and set $L_{\min} = 64$. The fit results suggest other corrections exist, so we add a sub-leading correction $y_2 = -1$. After adding this term, small‐size information is captured and $L_{\min}$ is reduced to 24. From the free $y_1$ fit, we choose a reasonable value for $y_1$ and then fix $y_1$ at its boundary to estimate $\eta$. Finally, we gradually increase $L_{\min}$ to test the stability and reliability of the fit results. For other $\sigma$ values, the fitting method is similar. Notably, for $\sigma = 15/8$, including only a $y_1$ correction term suffices and no $y_2$ term is needed. The reliability of the $S_2$ fit in Fig.~\ref{fig:S2_errorbar} illustrates this. Specific fit details are given in Table~\ref{fit:S_2}. For other quantities that estimate $\eta$ (e.g.  $C_1$, $\chi_k$), we apply the same fitting method, and their results agree closely with those from $S_2$. We list only their fit results (See Table~\ref{table:eta_all_obs_ensemble_EB}) without showing the detailed procedures.

We plot $S_2/L^2$ as a function of $L$ for various values of $\sigma$ in both ensembles, using double-logarithmic scales, as shown in Fig.~\ref{fig:S2_all}. According to Eq.\eqref{eq:S2_scaling}, the slope of these curves should asymptotically approach $-\eta$ at large system sizes. For reference, the dashed line represents the NN case with $\eta_{\rm SR}=5/24$. To facilitate a direct comparison of slopes across different $\sigma$, the data are rescaled by a constant factor $a$ so that they approximately align at $L=512$. The inset highlights results for smaller sizes ($L<512$), where finite-size corrections are pronounced. As shown in Fig.~\ref{fig:S2_all}, the data for all $\sigma$ values asymptotically follow a power-law scaling at sufficiently large $L$ and exhibit similar scaling behavior in both ensembles. For $\sigma=2.2$, the data converge to the NN reference curve at large $L$. In contrast, for $\sigma \leq 2$, the slopes clearly deviate from the reference line, with the discrepancy becoming stronger as $\sigma$ decreases.

\begin{figure}[t]
    \centering
    \includegraphics[width=\linewidth]{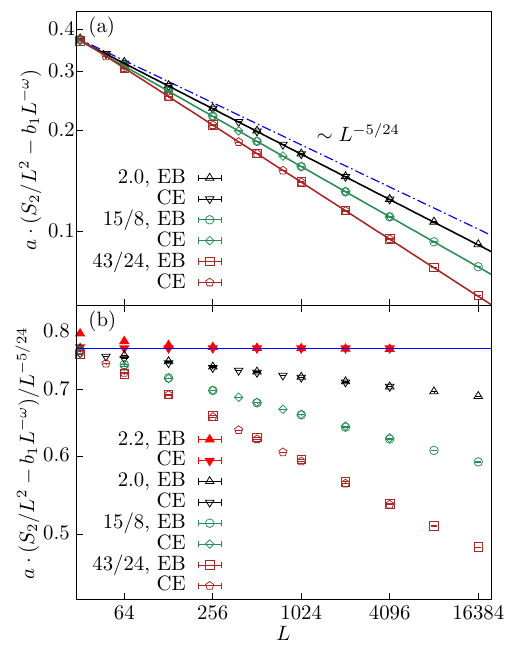}
    \caption{Panel (a) presents the finite-size scaling of $S_2/L^2$ for the CE and EB at various $\sigma$, after subtracting the leading correction term $L^{-\omega}$ with $\omega = \eta - y_1$. Rescaled by a constant factor, the data for both ensembles collapse across all $\sigma$ and display clear linear behavior in the log-log plot. The dashed line represents a guiding line with a slope of $\eta_{\text{SR}} = 5/24$, while the solid lines represent fits whose slopes are determined by the data. Panel (b) shows the same data divided by $L^{-5/24}$ to highlight deviations from the SR value, indicated by the blue horizontal line.}
    \label{fig:S2_Ensemble}
\end{figure}

The curves of $S_2/L^2$ in Fig.~\ref{fig:S2_all} are not perfectly linear even at the large $L$s, due to strong finite-size corrections. To further clarify the asymptotic behavior of $S_2$, we fit the data to Eq.\eqref{eq:S2_scaling}. As shown in Fig.~\ref{fig:S2_Ensemble}(a), after subtracting the contribution from the leading correction term $y_1$ and multiplying by a constant factor $a$, the curves display clear linearity, and data from the two ensembles collapse well onto each other, even down to $L=32$. For $\sigma = 2.0$, the extracted exponent is $\eta(2)=0.224(2)$, which exceeds the NN value $5/24 \approx 0.2083$ by about $0.016$. Further decreasing $\sigma$ yields a larger difference: for $\sigma = 43/24$, we find $\eta \approx 0.281(2)$, which is significantly larger than $\eta_{\rm SR}$. Fig.~\ref{fig:S2_Ensemble}(b) emphasizes these deviations by plotting $S_2/L^2$ divided by $L^{-5/24}$, which corresponds to the SR scaling relation. For $\sigma \leq 2$, the resulting curves display a nonzero slope on a log-log scale, and this slope becomes steeper as $\sigma$ decreases, highlighting that the system enters the LR universality when $\sigma \le 2$.

\begin{figure}[t]
    \centering
    \includegraphics[width=1\linewidth]{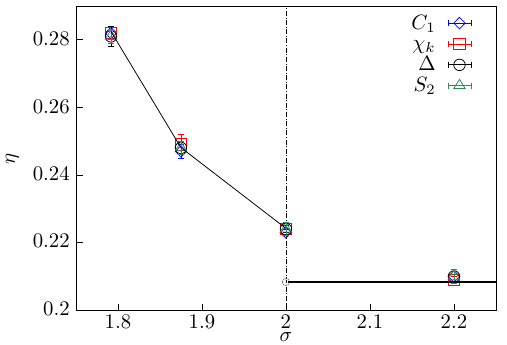}
    \caption{Anomalous dimension $\eta$ obtained independently from various observables as a function of $\sigma$. The estimates are mutually consistent within their respective uncertainties. At $\sigma = 2.2$, $\eta$ agrees closely with the SR value of $5/24$, as indicated by the horizontal line. However, for $\sigma \leq 2.0$, it deviates from the SR value, exhibiting a clear discontinuity at $\sigma = 2$. The deviation becomes more pronounced as $\sigma$ decreases further.} 
    \label{fig:eta_PD}
\end{figure}

The scaling analysis of $C_1$, $\chi_k$, and $\Delta$ yields values of $\eta$ that are fully consistent with those obtained from $S_2$, as summarized in Table~\ref{table:eta_all_obs_ensemble_EB} and shown in Fig.~\ref{fig:eta_PD}. In this paper, for different physical quantities yielding multiple estimates of the same critical exponent, we adopt the median of these estimates as the final value; when two medians occur, we use their mean, and similar for all other critical properties. Taken together, these results indicate that the LR–SR crossover occurs at $\sigma=2$ (within numerical accuracy), and that $\eta(\sigma)$ increases monotonically for smaller $\sigma$. This behavior is in clear contrast to the prediction of Ref.\cite{linderLongrangeEpidemicSpreading2008}, which, based on field-theoretical considerations, proposed that SR universality extends down to $\sigma=43/24$. Finally, the apparent discontinuity of $\eta(\sigma)$ at $\sigma=2$ should not be regarded as pathological; rather, it is consistent with previous analyses of dimensionless observables and arises naturally from the boundary case being in the LR universality class.


\begin{table}[ht]
    \caption{The anomalous dimension $\eta$ estimated from the various observables obtained in both EB and CE simulations.}
    \centering
    \begin{tabular}{c c | l l l l}
     \hline\hline
     $\sigma$ & & \multicolumn{1}{c}{$C_1$} & \multicolumn{1}{c}{$S_2$} & \multicolumn{1}{c}{$\chi_k$} & \multicolumn{1}{c}{$\Delta$} \\ \hline
     2.0      & EB &  0.223(1)    &  0.224(2)    & 0.224(2)   & 0.224(1) \\
              & CE &  0.224(1)    &  0.224(1)    & 0.225(3)   & \multicolumn{1}{c}{-} \\
     15/8     & EB &  0.247(2)    &  0.248(2)    & 0.249(3)   &  0.248(2)  \\
              & CE &  0.247(1)    &  0.247(2)    & 0.250(9)   & \multicolumn{1}{c}{-}  \\
     43/24    & EB &  0.282(2)    &  0.281(2)    & 0.282(2)   & 0.281(3) \\
              & CE &  0.282(1)    &  0.281(2)    & 0.28(1)    & \multicolumn{1}{c}{-}  \\
     \hline\hline
    \end{tabular}
    \label{table:eta_all_obs_ensemble_EB}
\end{table}

\subsection{Thermal Critical Exponents}


\begin{figure}[t]
    \centering
    \includegraphics[width=1\linewidth]{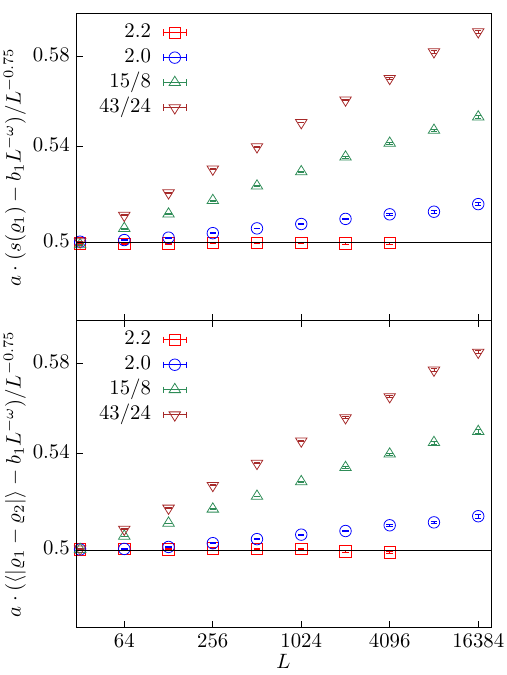}
    \caption{Log-log plot of (a) $s(\varrho_1)$ and (b) $\langle | \varrho_1-\varrho_2 | \rangle$ as functions of system size $L$ for various $\sigma$. After subtracting the leading correction $\omega = y_t - y_1$, the data are rescaled by a factor $a$ to collapse at $L=32$ and further divided by $L^{-0.75}$ to emphasize deviations from the SR value $0.75$. For $\sigma = 2.2$, the data align with the black horizontal line, while for $\sigma \leq 2$ they develop a clear slope, indicating that $y_t$ differs from its SR value.}
    \label{fig:yt2}
\end{figure}

\begin{figure}[t]
    \centering
    \includegraphics[width=1\linewidth]{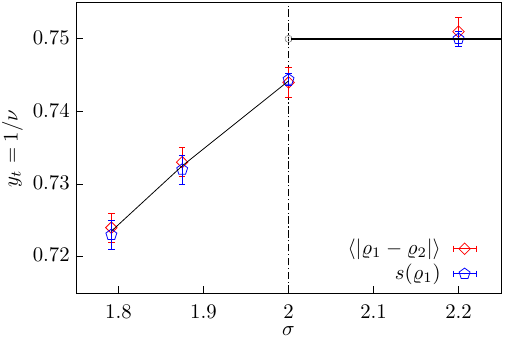}
    \caption{Thermal critical exponent $y_t \equiv1/\nu$ obtained from $s(\varrho_1)$ and $\langle|\varrho_1-\varrho_2|\rangle$ as a function of $\sigma$. The FSS analysis of the two observables yields consistent $y_t$ across all $\sigma$. At $\sigma = 2.2$, the value of $y_t$ agrees with the short-range value $0.75$ (solid horizontal line). At $\sigma = 2.0$, $y_t$ exhibits a discontinuous jump away from the SR value, and the deviation grows larger as $\sigma$ decreases further.}
    \label{fig:yt_vs_sigma}
\end{figure}

The thermal critical exponent is related to the correlation length critical exponent by $\nu \equiv 1/y_t$. Previous analysis of critical points and $R_p$ and $Q_m$ have produced preliminary estimates of $y_t$ for $\sigma = 2.2, 2.0, 15/8$ and $43/24$. To further characterize the behavior of $y_t$ in the crossover regime, we investigate the standard deviation of the dynamic pseudo-critical point $s(\varrho_1)$, and the averaged window between the first and second dynamic pseudo-critical points $\left< \left| \varrho _1-\varrho _2 \right| \right>$ in the EB. Both of these observables are expected to follow a FSS of form~\cite{li_explosive_2024}
\begin{align}
    \mathcal{O}_{L} \sim L^{-y_t}
\end{align}
To determine the value of the exponent $\nu$, we fit the data to the FSS ansatz
\begin{align}
    \mathcal{O}_{L} = L^{-y_t}(a+b_1L^{y_1}+b_2L^{y_2})
\end{align}
where $y_1$ and $y_2$ are the finite-size corrections. Similar to other critical-exponent fits, we attempt to add or remove sub-leading corrections of $-1$ or $-2$ to obtain the most reasonable fit ($\chi^2/\mathrm{DF}\approx1$), consistent with physical meaning. Since $s(\varrho_1)$ and $\langle|\varrho_1 - \varrho_2|\rangle$ give consistent results, we present only the fit details of $s(\varrho_1)$ in Table~\ref{fit:s}. We take the weighted average of $y_t$ from both quantities as our final estimate of $y_t$. Specifically, for $\sigma = 2$, our analysis gives $y_t =0.743(2)$ and for $\sigma = 43/24$, we find $y_t = 0.723(2)$. Note that the change in $y_t$ in the crossover regime is much smaller than that of $\eta$, making it very difficult to resolve in previous studies~\cite{grassbergerTwoDimensionalSIREpidemics2013, linderLongrangeEpidemicSpreading2008}. Similar issues also arise in our FSS analysis in CE, which is plagued by strong finite-size corrections. However, this small change of $y_t$ is accurately resolved by analyzing the FSS at the dynamic pseudo-critical point using large-scale simulations in the EB.

The FSS behaviors of $s(\varrho_1)$ and $\left< \left| \varrho _1-\varrho _2 \right| \right>$ are shown in Fig.~\ref{fig:yt2} to further validate our fitting results. By subtracting the leading correction and dividing by the scaling of the NN case, $L^{-1/\nu_{\rm SR}}$, Fig.~\ref{fig:yt2} enables a direct comparison of $y_t$ for various $\sigma$ cases with those of the SR universality. For both observables, similar scaling behavior can be seen: The rescaled data for $\sigma = 2.2$ converge to a horizontal line, confirming an exponent equal to $\nu_{\rm SR} = 3/4$; for $\sigma \le 2$, all data exhibit power-law scaling but with a finite slope, clearly deviating from the case of $\sigma > 2$. 

We plot $y_t$ extracted from both $s(\varrho_1)$ and $\left< \left| \varrho _1-\varrho _2 \right| \right>$ as a function of $\sigma$ in Fig.~\ref{fig:yt_vs_sigma}. It can be seen that $y_t$ obtained from both observables agrees with each other within the uncertainty, and the value of $y_t$ at $\sigma = 2$ is clearly different from the SR universality, manifesting a jump at the LR-SR crossover point. This find is again consistent with our previous results.

\subsection{Shortest-Path Exponent} 

The shortest‐path exponent $d_{\min}$ governs the scaling of the shortest path between two sites in a cluster as a function of Euclidean distance $r$ between these two sites~\cite{grassberger_numerical_1992, grassberger_spreading_1992},
\begin{align}
    \langle \ell(r)\rangle \sim r^{d_{\min}}.
\end{align}
Here, the shortest path $\ell$ corresponds to the minimum number of steps on a path of occupied sites in the cluster. The shortest-path exponent $d_{\min}$ of the 2D SR percolation universality class is estimated with high accuracy in Ref.~\cite{zhou_shortest-path_2012}, which gives $d_{\min,\rm SR} = 1.13077(2)$.

To determine the shortest path exponent $d_{\min}$, we study the FSS of the longest graph distance in the system, denoted as $S_1$~\cite{fang_backbone_2022}. The maximal graph distance of a cluster is measured using a BFS starting from a seed site in the cluster and finding the number of layers (depth) of the BFS. The longest graph distance is then the largest one among all clusters. At criticality, $S_1$ asymptotically diverges as
\begin{align}
    S_1 \sim L^{d_{\min}}.
\end{align}
We perform least-square fits to the FSS ansatz,
\begin{align}
    S_1 = L^{d_{\min}}(a+b_1L^{y_1}+b_2L^{y_2}),
\end{align}
where $y_1$ and $y_2$ are some finite-size corrections. During the fits, we find that two correction terms are required to describe the scaling of $S_1$. Fixing the sub-leading correction to $y_2 = -1$ yields the most stable fit. We then allow $y_1$ to vary in order to determine its reasonable range, and finally fix $y_1$ to obtain the final estimate of $d_{\min}$. For $\sigma = 2.2$, we obtain $d_{\rm min} = 1.130(1)$, consistent with the previous estimate of the SR case. For $\sigma = 2.0$, $15/8$, and $43/24$, we obtained $d_{\min} = 1.108(2)$, $1.073(3)$, and $1.058(3)$, respectively, as shown in Fig.~\ref{fig:d_min_vs_sigma}. The reliability of the $S_1$ fit is demonstrated in Appendix Fig.~\ref{fig:S1_errorbar}. Similar to other critical exponents, the shortest-path exponent exhibits a discontinuous jump at $\sigma = 2$, marking the crossover between LR and SR universality. In the LR regime, as $\sigma$ decreases, the $d_{\rm min}$ also decreases. Note that, unlike other exponents, the first derivative of the exponent appears to grow as $\sigma \to 2^-$, providing further evidence of a sharp change in universality at $\sigma = 2$. The details of the fits are summarized in Table~\ref{fit:dmin_CE}.

\begin{figure}[t]
    \centering
    \includegraphics[width=1\linewidth]{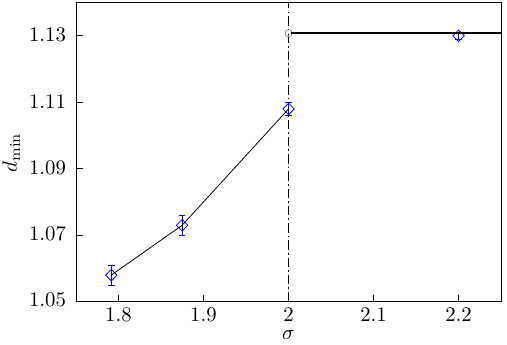}
    \caption{Shortest-path exponent $d_{\mathrm{min}}$ as a function of $\sigma$. At $\sigma = 2.2$, $d_{\mathrm{min}}$ agrees with the SR value $1.13077(2)$ (horizontal line). For $\sigma \leq 2$, it exhibits a discontinuous jump from the SR value, signaling the crossover from SR to LR universality, and decreases further as $\sigma$ decreases.}
    \label{fig:d_min_vs_sigma}
\end{figure}

\begin{figure}[t]
    \centering
    \includegraphics[width=1\linewidth]{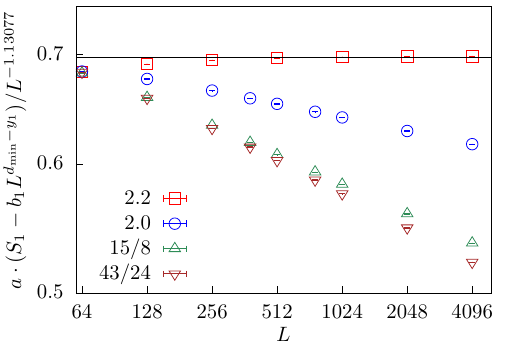}
    \caption{Finite-size scaling of $S_1$ for various $\sigma$, with the leading correction subtracted. To highlight deviations from SR universality, the data are divided by $L^{1.13077}$ and rescaled by a constant factor $a$. At $\sigma = 2.2$, the points converge to a horizontal line for $L \geq 512$, consistent with the NN case, whereas for $\sigma \leq 2$ the curves develop a clear slope, indicating a change in $d_{\min}$ and hence a different universality class.}
    \label{fig:smx}
\end{figure}

To further demonstrate the reliability of our estimates and compare the exponent across various $\sigma$, we subtract the contribution of the leading correction from $S_1$, divide it by $L^{1.13077}$, and plot the data as a function of $L$ in Fig.~\ref{fig:smx}. Due to substantial finite-size corrections, the data show curvature at small $L$, whereas for large system sizes, the data show a linear trend. For $\sigma = 2.2$, as $L$ increases, the curve gradually flattens, converging to the SR result. For $\sigma \le 2$, the data exhibits a clear slope, corresponding a $d_{\rm min}$ different from the SR case, and this slope grows larger as $\sigma$ decreases.

\section{Conclusion}
\label{sec:conclusion}

In this work, we conduct a detailed study of the LR bond percolation model using both the CE and the EB method. Through large-scale simulations and systematic FSS analysis, we estimate the universal values $R_p$ and $Q_m$ at criticality, as well as the critical exponents $\eta$, $y_t$, and $d_{\rm}$, for several representative $\sigma$s in the crossover regime. Our results provide direct evidence that the boundary between the LR and SR universality classes lies at $\sigma = 2.0$, rather than at $2 - \eta_{\mathrm{SR}}$~\cite{linderLongrangeEpidemicSpreading2008}. In particular, our work clearly demonstrates that the system falls under the LR universality class at $\sigma = 2$. As a result, a new picture of the SR-LR crossover emerges: the critical properties vary continuously within the LR regime for $\sigma \le 2$, but at $\sigma = 2$ they exhibit a discontinuous jump, producing a finite gap relative to the SR universality.




These results also shed light on the LR-SR crossover of the LR $O(n)$ model. In Ref.~\cite{picco2012}, it was also observed that the critical exponent $\eta$ at $\sigma = 2$ is slightly different from its SR values. Based on the new picture of the LR-SR crossover, this may not be a simple overestimation due to the finite-size effect, but a genuine discontinuity indicating a change in universality. Indeed, due to the subtlety of 2D Ising universality, this scenario requires further investigation. In addition, this result aligns with the 2D LR XY model~\cite{xiao_two-dimensional_2024, yao2025nonclassicalregimetwodimensionallongrange} and the 2D LR Heisenberg model~\cite{LRHeisenberg}, where the type or existence of the transition exhibits a sudden change at $\sigma = 2$. In all, our study brings a new perspective on understanding the SR-LR crossover and provides a new standpoint for future field theoretical studies. 


\acknowledgments
We acknowledge the support by the National Natural Science Foundation of China (NSFC) under Grant No. 12204173 and No. 12275263, as well as the Innovation Program for Quantum Science and Technology (under Grant No. 2021ZD0301900). YD is also supported by the Natural Science Foundation of Fujian Province 802 of China (Grant No. 2023J02032).
\\

\appendix
\section{Fitting result of universal ratios}

\begin{table}[t]
\caption{Summary of the universal values and critical points.}
\begin{adjustbox}{max width=\textwidth}
\begin{tabular}{l|llll}
\hline\hline
$\sigma$ &$\rho_c(R_p)$  &{$R_p$}       &$\rho_c(Q_m)$   & {$Q_m$}          \\\hline
NN       &      0.5       &\,\,0        & 0.5            & 0.870 48(5)      \\
2.2      & 0.454 409(1)   &\,\,0.000(1) & 0.454 409(2)   & 0.868 5(10)       \\\hline
2.0      & 0.436 506(2)   & -0.024(4)   & 0.436 505(2)   & 0.861(2)         \\
15/8     & 0.423 004(2)   & -0.076(6)   & 0.423 002(3)   & 0.843(3)         \\
43/24    & 0.413 054(2)   & -0.143(7)   & 0.413 050(5)   & 0.821(4)         \\\hline\hline
\end{tabular}
\end{adjustbox}
\normalsize
\label{table:fitting result}
\end{table}

In this section, we describe the detailed fitting process of the universal values of $R_p$ and $Q_m$ at critical points, using the FSS ansatz in Eq. (\ref{eq:fss_fitting_ansatz}). 

For $R_p$, we first perform unconstrained fits, keeping the polynomial terms with $m=2$ or $m=3$ and including both the $y_1$ and $y_2$ terms. To account for possible additional correction-to-scaling effects, we impose a lower cutoff $L \geq L_{\min}$ on the fitting data and systematically monitor how the residual $\chi^2$ value changes as $L_{\min}$ is increased. For a given ansatz, the preferred fit corresponds to the smallest $L_{\min}$ for which the goodness of fit remains reasonable and for which further increases in $L_{\min}$ do not lower the $\chi^2$ value by more than about one unit per degree of freedom. Here, ``reasonable" means $\chi^2/\mathrm{DF} \approx 1$, where DF denotes the number of degrees of freedom. In these unconstrained fits, $R_p$ shows only weak finite-size corrections, so the smallest cutoff $L_{\min}=8$ already produces stable results. Based on the unconstrained fit results, we then systematically examine the role of correction terms. For $\sigma = 2.2$, the result $y_2 = -1.5(2)$ are expressed using two correction exponents $y_2 = -1$ and $y_3 = -2$. The same approach applies to other values of $\sigma$. After adding and fixing the two sub-leading corrections, we fix $y_1$ within an appropriate range based on its estimate and refit the data. For $\sigma = 2.2$, we also fix $y_t$ to its SR value $0.75$. Including the $c_1$ term for $R_p$ only increases the fitting error without yielding further insight, so we set $c_1=0$. Through repeated fitting attempts, we obtain a precise estimate of $\rho_c$. For $\sigma = 15/8$ and $43/24$, the unconstrained fits give $y_2$ values very close to $-1$, so we also consider simplified fits with only one sub-leading correction fixed at $y_2=-1$. Combining all these results, we achieve high-precision estimates of both the critical points and the universal ratios, as summarized in Table~\ref{table:fitting result}, with detailed fit results presented in Table~\ref{fit:Rp}.

We apply the same fitting strategy to $Q_m$, but in this case, the finite-size corrections are much stronger. Consequently, small cutoffs cannot capture the correct scaling, and we set $L_{\min}=64$. Moreover, we find that including the $c_1$ term is necessary to obtain reliable estimates of $y_t$. Since the large cutoff $L_{\min}$ suppresses the ability to resolve higher-order corrections, we restrict our analysis to either a single correction with $y_2=-1$ or no additional correction. In particular, for $\sigma = 2.2$, fixing $y_t$ at $0.75$ gives stable fit results. For all $\sigma$, after fixing $y_1$ in a reasonable value obtained from the unconstrained fit, we systematically increase $L_{\min}$ to test fit stability and finally obtain the critical point $\rho_c$ and the universal ratio from $Q_m$, also shown in the Table~\ref{table:fitting result}. The fit details are in Table~\ref{table:Qm_fit}.

The results obtained from $R_p$ and $Q_m$ are consistent, and the corresponding estimates of $\rho_c$ agree within error bars. By taking both quantities into account, we obtain our final estimate of the critical point.





\begin{table*}[h]
\centering
\caption{Fitting results for the critical polynomial \(R_p\) under four distinct \(\sigma\) values (2.2, 2.0, 15/8, 43/24).}
\begin{adjustbox}{max width=\textwidth}
\begin{tabular*}{\textwidth}{@{\extracolsep{\fill}}ccclllllllllll}
\hline\hline
$\sigma$ & $L_{\min}$ & $\chi^2$/DF & $q_0$            & $q_1$    & $q_2$       & $y_t$    & $\rho_c$       & $y_1$    & $y_2$   & $y_3$ & $b_1$     & $b_2$    & $b_3$   \\ \hline
2.2      & 8          & 50.2/51     & -0.001(2)        & 1.72(3)  & -0.04(6)    & 0.745(3) & 0.454 408(1)   & 0.44(1)  & -1.5(2) & -     & -0.67(3)  & 0.7(1)   & -       \\
         & 8          & 53.2/52     & -0.001(2)        & 1.663(4) & -0.05(6)    & 0.75     & 0.454 408(2)   & 0.44(1)  & -1.5(2) & -     & -0.67(3)  & 0.7(1)   & -       \\
         & 24         & 33.5/33     & \,\,0.001(3)     & 1.660(4) & -0.04(6)    & 0.75     & 0.454 409(2)   & 0.43(2)  & -1      & -     & -0.66(6)  & 0.1(1)   & -       \\
         & 12         & 44.1/43     & \,\,0.000 6(13)  & 1.666(4) & -0.06(6)    & 0.75     & 0.454 409(1)   & 0.423(6) & -2      & -     & -0.632(8) & 1.4(1)   & -       \\
         & 16         & 38.6/38     & -0.001(2)        & 1.667(4) & -0.06(6)    & 0.75     & 0.454 408(1)   & 0.432(7) & -2      & -     & -0.65(1)  & 1.8(2)   & -       \\
         & 8          & 50.5/51     & -0.000 3(5)      & 1.663(4) & \,\,0.01(7) & 0.75     & 0.454 409 9(9) & 0.44     & -1      & -2    & -0.691(5) & 0.19(2)  & \,\,0.80(8) \\
         & 10         & 47.4/46     & -0.000 4(6)      & 1.663(4) & \,\,0.01(7) & 0.75     & 0.454 410(1)   & 0.44     & -1      & -2    & -0.690(7) & 0.18(3)  & \,\,0.8(2)  \\
         & 8          & 54.4/54     & \,\,0.000 4(5)   & 1.661(4) & -0.04(6)    & 0.75     & 0.454 409 4(8) & 0.43     & -1      & -2    & -0.659(5) & 0.10(2)  & \,\,0.97(8) \\
         & 10         & 47.7/48     & \,\,0.000 01(52) & 1.661(4) & -0.04(6)    & 0.75     & 0.454 409 0(8) & 0.43     & -1      & -2    & -0.656(6) & 0.08(3)  & \,\,1.1(1)  \\ \hline
2.0      & 8          & 42.2/43     & -0.020(7)        & 1.77(3)  & \,\,0.21(8)     & 0.736(2) & 0.436 507(2)   & 0.30(2)  & -1.2(2) & -     & -0.65(3)  & 0.48(4)  & -       \\
         & 8          & 43.4/43     & -0.031(2)        & 1.77(3)  & \,\,0.19(8)     & 0.736(2) & 0.436 504(1)   & 0.329(6) & -1      & -     & -0.720(8) & 0.47(2)  & -       \\
         & 10         & 37.8/37     & -0.029(3)        & 1.78(3)  & \,\,0.20(8)     & 0.735(2) & 0.436 504(2)   & 0.324(7) & -1      & -     & -0.71(1)  & 0.45(3)  & -       \\
         & 8          & 41.9/42     & -0.022(5)        & 1.77(3)  & \,\,0.21(8)     & 0.736(2) & 0.436 507(2)   & 0.31(1)  & -1      & -2    & -0.68(2)  & 0.32(8)  & \,\,0.4(2)  \\
         & 8          & 42.6/43     & -0.028 0(7)      & 1.77(3)  & \,\,0.19(8)     & 0.736(2) & 0.436 504 5(9) & 0.32     & -1      & -2    & -0.701(4) & 0.40(2)  & \,\,0.18(8) \\
         & 10         & 40.6/41     & -0.027 8(8)      & 1.77(3)  & \,\,0.20(8)     & 0.736(2) & 0.436 504 7(9) & 0.32     & -1      & -2    & -0.702(4) & 0.41(3)  & \,\,0.1(1)  \\
         & 8          & 42.1/42     & -0.019 8(7)      & 1.77(3)  & \,\,0.22(8)     & 0.736(2) & 0.436 507 4(9) & 0.3      & -1      & -2    & -0.666(3) & 0.28(2)  & \,\,0.43(8) \\
         & 10         & 38.5/38     & -0.020 0(8)      & 1.77(3)  & \,\,0.22(8)     & 0.736(2) & 0.436 507(1)   & 0.3      & -1      & -2    & -0.665(4) & 0.27(3)  & \,\,0.5(1)  \\ \hline
15/8     & 8          & 51.8/51     & -0.08(1)         & 1.73(3)  & \,\,0.51(8)     & 0.734(2) & 0.423 004(3)   & 0.24(2)  & -1.1(1) & -     & -0.61(2)  & 0.43(2)  & -       \\
         & 8          & 49.9/49     & -0.080(5)        & 1.71(3)  & \,\,0.5(1)      & 0.735(2) & 0.423 004(2)   & 0.246(6) & -1      & -     & -0.624(3) & 0.39(2)  & -       \\
         & 10         & 44.8/45     & -0.076(6)        & 1.71(3)  & \,\,0.6(1)      & 0.735(2) & 0.423 005(2)   & 0.241(7) & -1      & -     & -0.620(4) & 0.37(2)  & -       \\
         & 8          & 53.8/53     & -0.082 3(4)      & 1.73(3)  & \,\,0.49(8)     & 0.733(2) & 0.423 003 1(8) & 0.25     & -1      & -     & -0.627(1) & 0.405(4) & -       \\
         & 10         & 50.8/50     & -0.082 9(5)      & 1.73(3)  & \,\,0.48(8)     & 0.734(2) & 0.423 002 5(8) & 0.25     & -1      & -     & -0.625(2) & 0.396(5) & -       \\
         & 8          & 50.3/50     & -0.074 5(5)      & 1.71(3)  & \,\,0.52(7)     & 0.735(2) & 0.423 005 8(8) & 0.24     & -1      & -     & -0.621(1) & 0.376(4) & -       \\
         & 10         & 47.5/48     & -0.075 4(5)      & 1.72(3)  & \,\,0.49(7)     & 0.734(2) & 0.423 004 8(8) & 0.24     & -1      & -     & -0.618(2) & 0.365(5) & -       \\
         & 8          & 50.9/51     & -0.080(8)        & 1.72(3)  & \,\,0.51(8)     & 0.734(2) & 0.423 003(3)   & 0.24(1)  & -1      & -2    & -0.62(1)  & 0.35(6)  & \,\,0.2(2)  \\
         & 8          & 51.4/51     & -0.069 8(8)      & 1.73(3)  & \,\,0.52(8)     & 0.734(2) & 0.423 006(1)   & 0.23     & -1      & -2    & -0.604(3) & 0.26(2)  & \,\,0.43(7) \\
         & 10         & 47.0/47     & -0.069(1)        & 1.73(3)  & \,\,0.52(8)     & 0.734(2) & 0.423 006(1)   & 0.23     & -1      & -2    & -0.606(3) & 0.28(2)  & \,\,0.3(1)  \\
         & 8          & 52.2/52     & -0.083 9(8)      & 1.73(3)  & \,\,0.50(8)     & 0.733(2) & 0.423 002(1)   & 0.25     & -1      & -2    & -0.621(3) & 0.36(2)  & \,\,0.20(7) \\
         & 10         & 47.6/48     & -0.082 4(9)      & 1.72(3)  & \,\,0.53(8)     & 0.734(2) & 0.423 003(1)   & 0.25     & -1      & -2    & -0.627(3) & 0.41(2)  & -0.1(1) \\ \hline
43/24    & 8          & 41.3/41     & -0.14(2)         & 1.64(3)  & \,\,0.66(8)     & 0.735(2) & 0.413 054(4)   & 0.20(2)  & -1.1(1) & -     & -0.526(7) & 0.37(3)  & -       \\
         & 12         & 26.4/26     & -0.148(9)        & 1.65(3)  & \,\,0.66(8)     & 0.734(3) & 0.413 053(3)   & 0.21(1)  & -1      & -     & -0.536(3) & 0.33(3)  & -       \\
         & 16         & 26.3/26     & -0.142(1)        & 1.64(3)  & \,\,0.67(8)     & 0.735(3) & 0.413 054(1)   & 0.2      & -1      & -     & -0.535(3) & 0.31(1)  & -       \\
         & 24         & 24.4/23     & -0.142(1)        & 1.65(3)  & \,\,0.66(9)     & 0.734(3) & 0.413 054(2)   & 0.2      & -1      & -     & -0.534(4) & 0.31(2)  & -       \\
         & 16         & 25.3/25     & -0.151(1)        & 1.65(3)  & \,\,0.63(9)     & 0.734(3) & 0.413 052(1)   & 0.21     & -1      & -     & -0.537(3) & 0.34(1)  & -       \\
         & 24         & 22.9/22     & -0.150(1)        & 1.65(3)  & \,\,0.64(9)     & 0.734(3) & 0.413 052(2)   & 0.21     & -1      & -     & -0.538(4) & 0.35(2)  & -       \\
         & 8          & 40.1/41     & -0.14(1)         & 1.63(3)  & \,\,0.66(8)     & 0.735(2) & 0.413 054(3)   & 0.20(1)  & -1      & -2    & -0.533(5) & 0.29(6)  & \,\,0.2(2)  \\
         & 8          & 40.8/41     & -0.151(1)        & 1.64(3)  & \,\,0.64(8)     & 0.735(2) & 0.413 052(1)   & 0.21     & -1      & -2    & -0.536(3) & 0.33(2)  & \,\,0.10(8) \\ 
         & 10         & 27.9/28     & -0.151(2)        & 1.65(3)  & \,\,0.63(9)     & 0.734(3) & 0.413 052(2)   & 0.21     & -1      & -2    & -0.535(4) & 0.33(3)  & \,\,0.1(2)  \\ 
         & 8          & 41.4/41     & -0.133(1)        & 1.63(3)  & \,\,0.67(8)     & 0.735(2) & 0.413 056(1)   & 0.19     & -1      & -2    & -0.531(3) & 0.25(2)  & \,\,0.30(8) \\ 
         & 10         & 27.9/28     & -0.133(2)        & 1.64(3)  & \,\,0.66(8)     & 0.735(3) & 0.413 055(2)   & 0.19     & -1      & -2    & -0.528(4) & 0.23(3)  & \,\,0.4(2)  \\          
         \hline\hline
\end{tabular*}
\end{adjustbox}
\normalsize
\label{fit:Rp}
\end{table*}

\begin{table*}[ht]
\centering
\caption{Fitting results for the Binder ratio \(Q_m\) under four distinct \(\sigma\) values (2.2, 2.0, 15/8, 43/24).}
\begin{adjustbox}{max width=\textwidth}
\begin{tabular*}{\textwidth}{@{\extracolsep{\fill}}cccllllllllll}
\hline\hline
$\sigma$ & $L_{\min}$ & $\chi^2$/DF & $q_0$      & $q_1$    & $q_2$    & $y_t$    & $\rho_c$       & $y_1$    & $y_2$ & $b_1$      & $b_2$          & $c_1$   \\ \hline
2.2      & 128        & 16.1/17     & 0.868 7(5)  & 0.33(3)  & -0.27(5) & 0.749(9) & 0.454 406(1)   & 0.52(1)  & -     & -0.30(1)   & -              & 0.6(2)  \\
         & 128        & 15.4/16     & 0.868 5(5)  & 0.327(2) & -0.26(2) & 0.75     & 0.454 406(1)   & 0.52(1)  & -     & -0.30(2)   & -              & 0.63(7) \\
         & 64         & 25.7/25     & 0.869(1)   & 0.324(3) & -0.27(1) & 0.75     & 0.454 407(2)   & 0.48(5)  & -1    & -0.24(6)   & -0.2(2)        & 0.6(1)  \\
         & 64         & 26.4/26     & 0.870 6(2)  & 0.321(1) & -0.26(1) & 0.75     & 0.454 409 7(9) & 0.43     & -1    & -0.191(3)  & -0.30(2)       & 0.46(3) \\
         & 128        & 16.7/16     & 0.870 4(3)  & 0.322(2) & -0.27(2) & 0.75     & 0.454 409(1)   & 0.43     & -1    & -0.187(6)  & -0.34(6)       & 0.43(4) \\
         & 64         & 27.4/27     & 0.868 5(1)  & 0.322(1) & -0.27(1) & 0.75     & 0.454 405 6(8) & 0.52     & -1    & -0.297(4)  & -0.02(2)       & 0.66(5) \\
         & 128        & 15.9/16     & 0.868 6(3)  & 0.327(2) & -0.26(2) & 0.75     & 0.454 406(1)   & 0.52     & -1    & -0.30(1)   & \,\,0.01(8)    & 0.62(6) \\ \hline
2.0      & 64         & 26.2/25     & 0.861(2)   & 0.36(4)  & -0.28(5) & 0.74(1)  & 0.436 503(3)   & 0.38(3)  & -1    & -0.29(4)   & \,\,0.003(125) & 0.4(1)  \\
         & 64         & 25.1/25     & 0.863 4(3)  & 0.37(4)  & -0.29(6) & 0.74(1)  & 0.436 507(1)   & 0.35     & -1    & -0.253(2)  & -0.14(2)       & 0.4(1)  \\
         & 128        & 21.8/21     & 0.863 2(5)  & 0.38(5)  & -0.31(7) & 0.73(1)  & 0.436 506(1)   & 0.35     & -1    & -0.250(4)  & -0.17(5)       & 0.3(2)  \\
         & 64         & 26.8/26     & 0.859 1(2)  & 0.38(4)  & -0.30(5) & 0.73(1)  & 0.436 501(1)   & 0.41     & -1    & -0.321(3)  & \,\,0.10(2)    & 0.4(1)  \\
         & 128        & 21.3/21     & 0.859 5(4)  & 0.39(4)  & -0.32(6) & 0.73(1)  & 0.436 501(1)   & 0.41     & -1    & -0.327(6)  & \,\,0.15(6)    & 0.4(2)  \\ \hline
15/8     & 64         & 22.4/22     & 0.841(3)   & 0.44(6)  & -0.31(7) & 0.72(1)  & 0.422 999(4)   & 0.34(3)  & -1    & -0.30(3)   & \,\,0.1(1)     & 0.3(2)  \\
         & 64         & 23.6/23     & 0.845 5(3)  & 0.43(7)  & -0.29(8) & 0.72(1)  & 0.423 003(1)   & 0.31     & -1    & -0.267(2)  & -0.07(2)       & 0.3(2)  \\
         & 128        & 18.0/18     & 0.844 8(6)  & 0.33(9)  & -0.20(9) & 0.75(3)  & 0.423 001(2)   & 0.31     & -1    & -0.262(5)  & -0.14(6)       & 0.6(2)  \\
         & 64         & 23.0/23     & 0.838 9(3)  & 0.45(5)  & -0.32(6) & 0.72(1)  & 0.422 995(1)   & 0.37     & -1    & -0.330(3)  & \,\,0.17(2)    & 0.4(2)  \\
         & 128        & 17.8/18     & 0.839 2(5)  & 0.38(7)  & -0.24(7) & 0.74(2)  & 0.422 996(2)   & 0.37     & -1    & -0.333(6)  & \,\,0.20(6)    & 0.6(2)  \\ \hline
43/24    & 64         & 27.6/28     & 0.821(1)   & 0.585(9) & -0.34(4) & 0.700(2) & 0.413 050(2)   & 0.297(5) & -     & -0.266(2)  & -              & -       \\
         & 64         & 26.3/26     & 0.825 1(1)  & 0.588(9) & -0.33(4) & 0.699(2) & 0.413 056 0(8) & 0.28     & -     & -0.260 8(5) & -              & -       \\
         & 128        & 20.7/20     & 0.824 9(2)  & 0.59(1)  & -0.35(5) & 0.699(2) & 0.413 055 0(9) & 0.28     & -     & -0.260 1(8) & -              & -       \\
         & 64         & 28.2/28     & 0.820 5(1)  & 0.585(9) & -0.34(4) & 0.700(2) & 0.413 049 0(8) & 0.3      & -     & -0.267 8(5) & -              & -       \\
         & 128        & 23.8/24     & 0.820 7(2)  & 0.58(1)  & -0.33(4) & 0.700(2) & 0.413 050 0(9) & 0.3      & -     & -0.268 9(8) & -              & -       \\
         
         & 64         & 24.5/24     & 0.821(1)   & 0.46(8)  & -0.23(7) & 0.72(2)  & 0.413 049(2)   & 0.298(5) & -     & -0.267(2)  & -              & 0.3(2)  \\
         & 64         & 27.3/27     & 0.822 7(1)  & 0.49(7)  & -0.25(6) & 0.72(1)  & 0.413 052 6(8) & 0.29     & -     & -0.263 8(5) & -              & 0.2(2)  \\
         & 128        & 21.5/21     & 0.822 6(2)  & 0.51(9)  & -0.27(8) & 0.71(2)  & 0.413 052 3(9) & 0.29     & -     & -0.263 6(9) & -              & 0.2(2)  \\
         & 64         & 27.8/27     & 0.818 3(1)  & 0.45(6)  & -0.23(6) & 0.73(1)  & 0.413 045 2(8) & 0.31     & -     & -0.271 5(5) & -              & 0.3(1)  \\
         & 128        & 18.7/19     & 0.818 8(2)  & 0.5(1)   & -0.23(9) & 0.72(2)  & 0.413 047 1(9) & 0.31     & -     & -0.274(1)  & -              & 0.3(2)  \\

         & 64         & 26.0/26     & 0.822 2(4)  & 0.47(7)  & -0.24(6) & 0.72(1)  & 0.413 051(1)   & 0.29     & -1    & -0.261(2)  & -0.02(2)       & 0.3(2)  \\
         & 128        & 18.8/19     & 0.820 7(7)  & 0.4(1)   & -0.22(9) & 0.73(2)  & 0.413 047(2)   & 0.29     & -1    & -0.251(4)  & -0.16(5)       & 0.2(2)  \\
         & 64         & 22.2/22     & 0.819 3(4)  & 0.47(7)  & -0.24(7) & 0.72(1)  & 0.413 048(1)   & 0.31     & -1    & -0.278(2)  & \,\,0.05(2)    & 0.3(1)  \\
         & 128        & 17.9/18     & 0.818 2(7)  & 0.4(1)   & -0.22(9) & 0.73(2)  & 0.413 045(2)   & 0.31     & -1    & -0.269(5)  & -0.06(6)       & 0.4(2)  \\\hline\hline
\end{tabular*}
\end{adjustbox}
\normalsize
\label{table:Qm_fit}
\end{table*}

\begin{table*}[h]
\centering
\caption{Fitting results for the  \(S_2\) under four distinct \(\sigma\) values in the EB(2.2, 2.0, 15/8, and 43/24).}
\begin{adjustbox}{max width=\textwidth}
\begin{tabular*}{0.9\textwidth}{@{\extracolsep{\fill}}cccllllll}
\hline\hline
$\sigma$ & $L_{\min}$ & $\chi^2$/DF & $\eta$    & $a$       & $b_1$    & $b_2$        & $y_1$    & $y_2$ \\\hline
2.2      & 64         & 1.9/3       & 0.211 4(7) & 0.481(3)  & 0.69(3)  & -            & -0.59(2)  & -     \\
         & 24         & 2.2/4       & 0.210(1)  & 0.475(7)  & 0.50(9)  & \,\,0.4(2)   & -0.52(5)  & -1    \\
         & 24         & 3.0/5       & 0.208 3(3) & 0.464(1)  & 0.410(7) & \,\,0.56(2)  & -0.47     & -1    \\
         & 32         & 2.7/4       & 0.208 4(4) & 0.465(2)  & 0.407(9) & \,\,0.58(3)  & -0.47     & -1    \\
         & 24         & 2.7/5       & 0.211 5(2) & 0.481(1)  & 0.61(1)  & \,\,0.20(3)  & -0.57     & -1    \\
         & 32         & 2.3/4       & 0.211 4(3) & 0.480(1)  & 0.62(1)  & \,\,0.18(4)  & -0.57     & -1    \\\hline
2.0      & 128        & 3.2/4       & 0.224(1)  & 0.400(7)  & 0.74(3)  & -            & -0.46(2)  & -     \\
         & 32         & 3.7/5       & 0.224(2)  & 0.398(9)  & 0.69(6)  & \,\,0.2(1)   & -0.45(3)  & -1    \\
         & 32         & 4.3/6       & 0.221 7(4) & 0.389(1)  & 0.636(6) & \,\,0.36(2)  & -0.42     & -1    \\
         & 64         & 3.3/5       & 0.222 2(5) & 0.391(2)  & 0.63(1)  & \,\,0.42(5)  & -0.42     & -1    \\
         & 32         & 4.9/6       & 0.225 8(4) & 0.408(1)  & 0.779(8) & \,\,0.02(3)  & -0.48     & -1    \\
         & 64         & 4.6/5       & 0.225 5(5) & 0.407(2)  & 0.79(1)  & -0.02(7)     & -0.48     & -1    \\\hline
15/8     & 64         & 7.4/5       & 0.248(2)  & 0.369(7)  & 0.84(1)  & -            & -0.417(9) & -     \\
         & 64         & 8.6/6       & 0.246 2(4) & 0.364(1)  & 0.829(3) & -            & -0.41     & -     \\
         & 128        & 5.1/5       & 0.246 7(5) & 0.366(2)  & 0.823(4) & -            & -0.41     & -     \\
         & 64         & 10.9/6      & 0.249 9(4) & 0.380(1)  & 0.855(4) & -            & -0.43     & -     \\
         & 128        & 7.4/5       & 0.249 7(6) & 0.379(2)  & 0.857(6) & -            & -0.43     & -     \\ \hline
43/24    & 24         & 7.0/6       & 0.281(2)  & 0.39(1)   & 0.93(5)  & -0.1(1)      & -0.41(2)  & -1    \\
         & 24         & 8.2/7       & 0.283 2(5) & 0.404(2)  & 0.977(7) & -0.24(2)     & -0.43     & -1    \\
         & 32         & 7.9/6       & 0.283 3(6) & 0.404(2)  & 0.975(9) & -0.23(4)     & -0.43     & -1    \\
         & 32         & 4.7/6       & 0.278 2(5) & 0.380(2)  & 0.865(6) & \,\,0.07(2)  & -0.39     & -1    \\
         & 64         & 4.0/5       & 0.278 7(7) & 0.382(3)  & 0.86(1)  & \,\,0.12(6)  & -0.39     & -1    \\ \hline\hline
\end{tabular*}
\end{adjustbox}
\normalsize
\label{fit:S_2}
\end{table*}

\begin{figure*}[h]
    \centering
    \includegraphics[width=\linewidth]{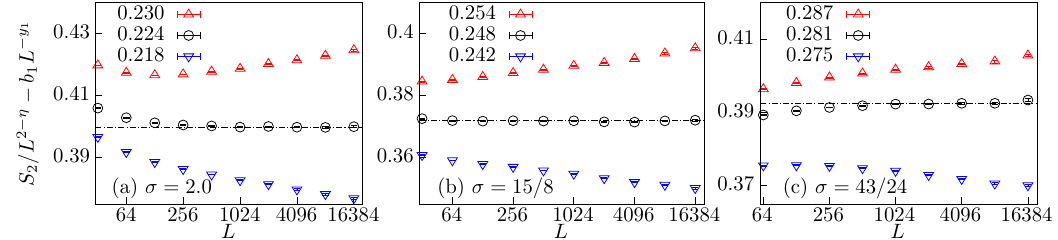}
    \caption{Demonstration of the reliability of the fit of $S_2$.The red data (blue data) is the result of fitting the center value (black) plus (minus) 3 times the error bar.}
    \label{fig:S2_errorbar}
\end{figure*}

\begin{table*}[h]
\centering
\caption{Fitting results for the  \(\rho_1\) under four distinct \(\sigma\) values (2.2, 2.0, 15/8, and 43/24).}
\begin{adjustbox}{max width=\textwidth}
\begin{tabular*}{0.9\textwidth}{@{\extracolsep{\fill}}ccclllllllll}
\hline\hline
$\sigma$ & $L_{\min}$ & $\chi^2$/DF & $y_t$    & $\rho_c$     & $a$        & $b_1$    & $b_2$    & $b_3$    & $y_1$   & $y_2$ & $y_3$ \\ \hline
2.2      & 12         & 8.2/5       & 0.75     & 0.454 407(4) & -0.442(6)  & 0.7(1)   & -0.4(3)  & -1.9(9)  & -0.42(4) & -1    & -2    \\
         & 12         & 9.5/6       & 0.75     & 0.454 409(2) & -0.448 1(9) & 0.608(8) & -0.08(5) & -2.7(3)  & -0.38    & -1    & -2    \\
         & 16         & 9.3/5       & 0.75     & 0.454 410(2) & -0.448(1)  & 0.61(1)  & -0.06(8) & -2.9(6)  & -0.38    & -1    & -2    \\
         & 12         & 9.7/6       & 0.75     & 0.454 403(1) & -0.436 9(8) & 0.84(1)  & -0.72(6) & -1.1(3)  & -0.46    & -1    & -2    \\
         & 16         & 7.9/5       & 0.75     & 0.454 404(2) & -0.437 4(9) & 0.85(1)  & -0.79(9) & -0.5(6)  & -0.46    & -1    & -2    \\ \hline
2.0      & 8          & 11.9/7      & 0.72(2)  & 0.436 506(5) & -0.33(6)   & 0.7(2)   & -0.7(4)  & -1.2(5)  & -0.4(1)  & -1    & -2    \\
         & 8          & 13.5/8      & 0.744(3) & 0.436 501(2) & -0.420(9)  & 0.63(2)  & -0.41(6) & -1.5(2)  & -0.3     & -1    & -2    \\
         & 10         & 10.8/7      & 0.742(3) & 0.436 503(3) & -0.41(2)   & 0.61(3)  & -0.33(9) & -1.9(3)  & -0.3     & -1    & -2    \\
         & 8          & 12.4/8      & 0.715(2) & 0.436 509(2) & -0.304(5)  & 0.82(3)  & -0.93(8) & -0.9(2)  & -0.5     & -1    & -2    \\
         & 10         & 12.2/7      & 0.714(3) & 0.436 509(2) & -0.304(6)  & 0.81(4)  & -0.9(1)  & -1.0(3)  & -0.5     & -1    & -2    \\ \hline
15/8    & 12         & 0.9/6       & 0.72(2)  & 0.423 007(6) & -0.30(7)   & 0.44(3)  & -0.1(1)  & -2.7(3)  & -0.27(8) & -1    & -2    \\
         & 8          & 8.2/9       & 0.744(2) & 0.423 005(2) & -0.431(9)  & 0.56(2)  & -0.18(4) & -2.1(1)  & -0.19    & -1    & -2    \\
         & 10         & 3.7/8       & 0.740(2) & 0.423 007(1) & -0.416(8)  & 0.53(2)  & -0.09(4) & -2.5(2)  & -0.19    & -1    & -2    \\
         & 8          & 3.9/9       & 0.706(2) & 0.423 011(1) & -0.258(3)  & 0.48(1)  & -0.34(3) & -1.80(8) & -0.35    & -1    & -2    \\
         & 10         & 2.2/8       & 0.704(1) & 0.423 012(1) & -0.253(3)  & 0.46(1)  & -0.28(3) & -2.0(1)  & -0.35    & -1    & -2    \\ \hline
43/24   & 10         & 6.7/7       & 0.68(2)  & 0.413 058(5) & -0.16(4)   & 0.4(1)   & -0.4(3)  & -1.8(5)  & -0.4(2)  & -1    & -2    \\
         & 10         & 8.4/8       & 0.732(4) & 0.413 051(2) & -0.31(1)   & 0.46(2)  & -0.17(6) & -2.4(2)  & -0.2     & -1    & -2    \\
         & 12         & 5.9/7       & 0.728(4) & 0.413 052(2) & -0.30(1)   & 0.43(2)  & -0.08(7) & -2.8(3)  & -0.2     & -1    & -2    \\
         & 10         & 8.2/8       & 0.659(3) & 0.413 064(2) & -0.132(3)  & 0.73(4)  & -0.9(1)  & -1.2(2)  & -0.6     & -1    & -2    \\
         & 12         & 8.1/7       & 0.659(4) & 0.413 064(2) & -0.132(3)  & 0.73(5)  & -0.9(1)  & -1.2(4)  & -0.6     & -1    & -2    \\ \hline\hline
\end{tabular*}
\end{adjustbox}
\normalsize
\label{fit:rho_1}
\end{table*}

\begin{table*}[t]
\centering
\caption{Fitting results for the  \(s(\varrho_1)\) under four distinct \(\sigma\) values in the EB(2.2, 2.0, 15/8, and 43/24).}
\begin{adjustbox}{max width=\textwidth}
\begin{tabular*}{0.9\textwidth}{@{\extracolsep{\fill}}cccllllll}
\hline\hline
$\sigma$ & $L_{\min}$ & $\chi^2$/DF & $y_t$     & $a$       & $b_1$     & $b_2$     & $y_1$    & $y_2$ \\\hline
2.2      & 24         & 1.1/4       & 0.750(1)  & 0.504(8)  & -0.48(8)  & 0.2(2)    & -0.51(5)  & -1    \\
         & 24         & 1.4/5       & 0.748 6(2) & 0.498(1)  & -0.56(1)  & 0.33(2)   & -0.56     & -1    \\
         & 32         & 0.8/4       & 0.748 8(2) & 0.499(1)  & -0.57(1)  & 0.36(3)   & -0.56     & -1    \\
         & 24         & 1.5/5       & 0.751 3(3) & 0.513(1)  & -0.405(7) & 0.04(2)   & -0.46     & -1    \\
         & 32         & 1.5/4       & 0.751 3(4) & 0.513(2)  & -0.40(1)  & 0.04(3)   & -0.46     & -1    \\\hline
2.0      & 12         & 9.6/8       & 0.744 4(8) & 0.509(4)  & -0.325(2) & -         & -0.380(9) & -     \\
         & 12         & 11.2/9      & 0.743 6(2) & 0.504 2(7) & -0.324(2) & -         & -0.39     & -     \\
         & 16         & 11.1/8      & 0.743 6(2) & 0.504 2(9) & -0.323(2) & -         & -0.39     & -     \\
         & 12         & 11.0/9      & 0.745 3(2) & 0.513 6(8) & -0.327(2) & -         & -0.37     & -     \\
         & 16         & 8.7/8       & 0.745 1(2) & 0.512 9(9) & -0.325(2) & -         & -0.37     & -     \\
         \hline
15/8     & 12         & 8.6/7       & 0.733(1)  & 0.470(8)  & -0.245(4) & -0.5(1)   & -0.38(3)  & -2    \\
         & 12         & 9.9/8       & 0.734 2(3) & 0.479(1)  & -0.244(3) & -0.63(6)  & -0.35     & -2    \\
         & 16         & 6.7/7       & 0.733 9(3) & 0.478(1)  & -0.240(3) & -0.8(1)   & -0.35     & -2    \\
         & 12         & 10.3/8      & 0.731 0(3) & 0.462(1)  & -0.249(3) & -0.35(7)  & -0.41     & -2    \\
         & 16         & 11.3/8      & 0.731 0(4) & 0.462(1)  & -0.249(4) & -0.4(1)   & -0.41     & -2    \\\hline
43/24    & 12         & 10.3/7      & 0.723(2)  & 0.437(8)  & -0.201(7) & -0.5(2)   & -0.42(4)  & -1     \\
         & 12         & 10.9/7      & 0.724 7(4) & 0.444(1)  & -0.195(4) & -0.65(6)  & -0.38     & -1    \\ 
         & 16         & 10.2/6      & 0.724 6(5) & 0.443(2)  & -0.193(4) & -0.7(1)   & -0.38     & -1    \\ 
         & 12         & 9.8/7       & 0.721 6(3) & 0.429(1)  & -0.207(3) & -0.37(6)  & -0.46     & -1    \\ 
         & 16         & 9.5/6       & 0.721 7(4) & 0.429(1)  & -0.208(5) & -0.3(1)   & -0.46     & -1    \\ 
         \hline\hline
\end{tabular*}
\end{adjustbox}
\normalsize
\label{fit:s}
\end{table*}

\begin{figure*}[h]
    \centering
    \includegraphics[width=\linewidth]{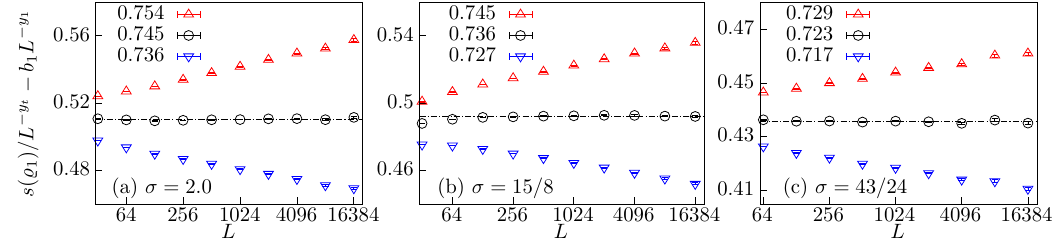}
    \caption{Demonstration of the reliability of the fit of $s(\varrho_1)$.The red data (blue data) is the result of fitting the center value (black) plus (minus) 3 times the error bar}
    \label{fig:sig_errorbar}
\end{figure*}

\begin{table*}[h]
\centering
\caption{Fitting results for the  \(d_{\min}\) under four distinct \(\sigma\) values in the CE( 2.2, 2.0, 15/8, and 43/24).}
\begin{adjustbox}{max width=\textwidth}
\begin{tabular*}{0.9\textwidth}{@{\extracolsep{\fill}}cccllllll}
\hline\hline
$\sigma$ & $L_{\min}$ & $\chi^2$/DF & $d_{\min}$ & $a$      & $b_1$    & $b_2$     & $y_1$     & $y_2$ \\ \hline
2.2      & 24        & 1.1/3        & 1.130(1)   & 0.702(8) & 1.45(3)  & -1.08(7)  & -0.416(8)  & -1    \\
         & 24        & 4.8/4        & 1.130 8(3)  & 0.699(2) & 1.421(7) & -1.00(2)  & -0.41      & -1    \\
         & 32        & 4.8/3        & 1.130 8(5)  & 0.699(3) & 1.42(1)  & -1.00(3)  & -0.41      & -1    \\
         & 24        & 6.9/5        & 1.129 6(3)  & 0.709(2) & 1.460(8) & -1.10(2)  & -0.42      & -1    \\
         & 32        & 6.7/4        & 1.129 7(4)  & 0.708(3) & 1.46(1)  & -1.11(3)  & -0.42      & -1    \\
         \hline
2.0      & 12        & 7.8/9        & 1.109(1)   & 0.64(1)  & 1.466(6) & -1.20(3)  & -0.351(4)  & -1    \\
         & 12        & 12.0/10      & 1.106 6(3)  & 0.663(2) & 1.480(4) & -1.269(6) & -0.36      & -1    \\
         & 16        & 12.0/9       & 1.106 6(3)  & 0.663(2) & 1.480(4) & -1.27(1)  & -0.36      & -1    \\
         & 8         & 12.7/11      & 1.109 4(3)  & 0.641(2) & 1.461(3) & -1.182(4) & -0.35      & -1    \\\hline
15/8     & 64        & 0.4/4        & 1.076(2)   & 0.71(2)  & 1.37(2)  & -1.3(1)   & -0.316(9)  & -1    \\
         & 24        & 8.1/7        & 1.070 8(4)  & 0.750(3) & 1.372(7) & -1.38(2)  & -0.33      & -1    \\
         & 32        & 4.4/6        & 1.071 3(4)  & 0.746(3) & 1.381(7) & -1.41(2)  & -0.33      & -1    \\
         & 24        & 1.4/7        & 1.076 3(2)  & 0.698(2) & 1.352(3) & -1.197(7) & -0.31      & -1    \\
         & 32        & 0.9/6        & 1.076 5(2)  & 0.697(2) & 1.355(3) & -1.209(8) & -0.31      & -1    \\
\hline
43/24    & 10        & 8.7/10       & 1.058(2)   & 0.67(2)  & 1.26(1)  & -1.00(2)  & -0.254(4)  & -1    \\
         & 12        & 6.1/9        & 1.056(2)   & 0.69(2)  & 1.25(1)  & -1.03(2)  & -0.258(4)  & -1    \\
         & 10        & 10.2/11      & 1.060 6(5)  & 0.644(4) & 1.277(5) & -0.971(4) & -0.25      & -1    \\
         & 12        & 9.7/10       & 1.060 8(6)  & 0.643(4) & 1.279(5) & -0.973(5) & -0.25      & -1    \\
         & 16        & 9.7/9        & 1.060 8(7)  & 0.643(5) & 1.279(6) & -0.973(7) & -0.25      & -1    \\ 
         & 12        & 6.2/10       & 1.054 8(4)  & 0.701(3) & 1.247(4) & -1.036(4) & -0.26      & -1    \\
         & 16        & 4.0/9        & 1.055 1(4)  & 0.698(3) & 1.252(4) & -1.045(5) & -0.26      & -1    \\          
         \hline\hline
\end{tabular*}
\end{adjustbox}
\normalsize
\label{fit:dmin_CE}
\end{table*}

\begin{figure*}[h]
    \centering
    \includegraphics[width=\linewidth]{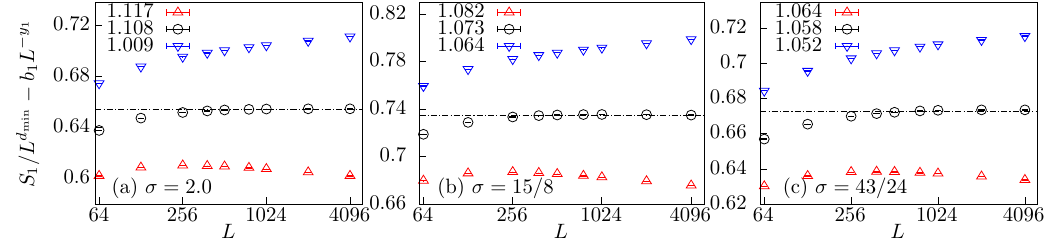}
    \caption{Demonstration of the reliability of the fit of $S_1$.The red data (blue data) is the result of fitting the center value (black) plus (minus) 3 times the error bar}
    \label{fig:S1_errorbar}
\end{figure*}

\bibliography{ref}

\end{document}